\def\final{0}

\documentclass[superscriptaddress, twocolumn,
notitlepage, amsmath, amssymb, aps,  showkeys,
pra,
10pt, letterpaper
]{revtex4-2}

\usepackage{float}

\usepackage{graphicx}
\usepackage{dcolumn}
\usepackage{bm}

\usepackage{fancyhdr}
\usepackage{latexsym}
\usepackage{graphicx,color}
\usepackage{grffile}		
\usepackage[pdfstartview=FitH,hidelinks]{hyperref}
\usepackage{xargs}
\usepackage{shadow,epsf,amsthm,amssymb,amsmath}

\usepackage{enumerate}

\usepackage{setspace}                           
 
\usepackage{lipsum}

\usepackage{tikz}	
\usetikzlibrary{backgrounds,fit,decorations.pathreplacing}  

\usepackage[colorinlistoftodos,prependcaption]{todonotes}

\usepackage{subcaption}

\usepackage[ruled,vlined,noresetcount]{algorithm2e}
\SetKwInput{KwOutput}{Output}              
\SetKwInOut{Input}{Input}
\SetKwInput{Output}{Output} 

\DeclareMathOperator*{\argmin}{arg\,min}
\DeclareMathOperator*{\argmax}{arg\,max}

\newtheorem{definitionenv}{Definition}
\newtheorem{lemmaenv}[definitionenv]{Lemma}
\newtheorem{theoremenv}[definitionenv]{Theorem}
\newtheorem{corollaryenv}[definitionenv]{Corollary}
\newtheorem{propositionenv}[definitionenv]{Proposition}
\newtheorem{conjectureenv}[definitionenv]{Conjecture}
\newtheorem{remarkenv}[definitionenv]{Remark}
\newenvironment{remark}{\begin{remarkenv}\rm}{\end{remarkenv}}
\newcommand{\br}{\begin{remark}}
\newcommand{\er}{\end{remark}}

\newtheorem{exampleenv}{Example}
\newtheorem{app-lemmaenv}[section]{Lemma}

\newenvironment{definition}{\begin{definitionenv}\rm}{\end{definitionenv}}
\newenvironment{lemma}{\begin{lemmaenv}\rm}{\end{lemmaenv}}
\newenvironment{theorem}{\begin{theoremenv}\rm}{\end{theoremenv}}
\newenvironment{corollary}{\begin{corollaryenv}\rm}{\end{corollaryenv}}
\newenvironment{example}{\begin{exampleenv}\rm}{\end{exampleenv}}
\newenvironment{proposition}{\begin{propositionenv}\rm}{\end{propositionenv}}
\newenvironment{conjecture}{\begin{conjectureenv}\rm}{\end{conjectureenv}}
\newenvironment{app-lemma}{\begin{app-lemmaenv}\rm}{\end{app-lemmaenv}}

\newcommand{\bd}{\begin{definition}}
\newcommand{\ed}{\end{definition}}
\newcommand{\bl}{\begin{lemma}}
\newcommand{\el}{\end{lemma}}
\newcommand{\elp}{\hspace*{\fill} $\Box$
                 \end{lemma}}
\newcommand{\bt}{\begin{theorem}}
\newcommand{\et}{\end{theorem}}
\newcommand{\etp}{\hspace*{\fill} $\Box$
                 \end{theorem}}
\newcommand{\bc}{\begin{corollary}}
\newcommand{\ec}{\end{corollary}}
\newcommand{\ecp}{\hspace*{\fill} $\Box$
                 \end{corollary}}
\newcommand{\bcj}{\begin{conjecture}}
\newcommand{\ecj}{\end{conjecture}}

\newcommand{\be}{\begin{example}}
\newcommand{\ee}{\end{example}}
\newcommand{\eep}{\hspace*{\fill} $\Box$
                 \end{example}}
\newcommand{\bp}{\begin{proposition}}
\newcommand{\ep}{\end{proposition}}
\newcommand{\epp}{
                 \end{proposition}}

\newcommand{\bra}[1]{\langle#1|}
\newcommand{\ket}[1]{|#1\rangle}

\newcommand{\wt}[1]{\text{wt}\left(#1\right)}
\newcommand{\gw}[1]{\text{gw}\left(#1\right)}

\newcommand{\eeq}{ \setcounter{equation} {\value{enumi}}}

\newcommand{\hC}{\hat{C}}
\newcommand{\hB}{\hat{B}}
\newcommand{\hX}{\hat{X}}
\newcommand{\hZ}{\hat{Z}}
\newcommand{\hI}{\hat{I}}
\newcommand{\hY}{\hat{Y}}

\newcommand{\cC}{\mathcal{C}}

\newcommand{\cG}{\mathcal{G}}

\newcommand{\cS}{\mathcal{S}}

\newcommand{\bma}{{\bm a}}
\newcommand{\bmb}{{\bm b}}

\newcommand{\bme}{{\bm e}}
\newcommand{\bms}{{\bm s}}

\newcommand{\bmu}{{\bm u}}
\newcommand{\bmv}{{\bm v}}
\newcommand{\bmx}{{\bm x}}
\newcommand{\bmy}{{\bm y}}
\newcommand{\bmz}{{\bm z}}

\def\beq{\begin{equation}}
\def\eeq{\end{equation}}

\def\bean{\begin{IEEEeqnarray*}{rCl}}
\def\eean{\end{IEEEeqnarray*}}

\ifnum\final=1
\newcommand{\mynote}[2]{{\color{#1} \marginpar{\tiny #2}}}
\newcommand{\mybignote}[2]{{\color{#1} $\langle \langle$ #2$\rangle \rangle$}}
\newcommandx{\rednote}[2][1=]{\todo[linecolor=red,backgroundcolor=red!25,bordercolor=red,#1]{#2}}
\newcommandx{\bluenote}[2][1=]{\todo[linecolor=blue,backgroundcolor=blue!25,bordercolor=blue,#1]{#2}}
\newcommandx{\yellownote}[2][1=]{\todo[linecolor=yellow,backgroundcolor=yellow!25,bordercolor=yellow,#1]{#2}}
\newcommandx{\greennote}[2][1=]{\todo[inline,linecolor=olive,backgroundcolor=green!25,bordercolor=olive,#1]{#2}}

\newcommand{\rmark}[1]{{\color{red} #1}}
\newcommand{\bmark}[1]{{\color{blue} #1}}

\else
\newcommand{\mynote}[2]{}
\newcommand{\mybignote}[2]{}
\newcommand{\rednote}[2][1=]{}
\newcommand{\bluenote}[2][1=]{}
\newcommand{\greennote}[2][1=]{}
\newcommand{\yellownote}[2][1=]{}
\newcommand{\rmark}[1]{#1}
\newcommand{\bmark}[1]{#1}

\fi
\newcommand{\eq}[1]{Eq.\,(\ref{#1})}

\newcommand{\MaxCut}{{\sf MaxCut}}
\newcommand{\gE}{{\sf E}}
\newcommand{\gG}{{\sf G}}
\newcommand{\gV}{{\sf V}}

\newenvironment{smallarray}[1]
 {\null\,\vcenter\bgroup\scriptsize
  \arraycolsep=.13885em
  \hbox\bgroup$\array{@{}#1@{}}}
 {\endarray$\egroup\egroup\,\null}

\DeclareMathOperator{\avg}{avg}

\begin{document}

\preprint{APS/123-QED}

\title{Syndrome decoding by quantum approximate optimization
}

\author{Ching-Yi Lai}
\email{cylai@nycu.edu.tw}
\affiliation{\footnotesize Institute of Communications Engineering, National Yang Ming Chiao Tung University, Hsinchu 30010, Taiwan}
\affiliation{\footnotesize
 Physics Division, National Center for Theoretical Sciences, Taipei 10617, Taiwan 
}

\author{Kao-Yueh Kuo}
\email{kywukuo@gmail.com}
\affiliation{\footnotesize School of Mathematical and Physical Sciences, University of Sheffield, Sheffield S3 7RH, United Kingdom}

\author{Bo-Jyun Liao}
\affiliation{\footnotesize Institute of Communications Engineering, National Yang Ming Chiao Tung University, Hsinchu 30010, Taiwan}

\begin{abstract}
The syndrome decoding problem is known to be NP-complete. 
The goal of the decoder is to find an error of low weight that corresponds to a given syndrome obtained from a parity-check matrix.  We use the quantum approximate optimization algorithm (QAOA) to address the syndrome decoding problem with elegantly-designed  reward Hamiltonians based on both generator and check matrices for classical and quantum codes.
We evaluate the level-4 check-based QAOA decoding of the [7,4,3] Hamming code, as well as the level-4 generator-based QAOA decoding of the [[5,1,3]] quantum code.
Remarkably, the simulation  results demonstrate that the decoding performances  match those of the maximum likelihood decoding. 
Moreover,  we explore the possibility of enhancing QAOA  by introducing additional redundant clauses to a combinatorial optimization problem while keeping the number of qubits unchanged. 
Finally, we study QAOA decoding of degenerate quantum codes.
Typically, conventional decoders aim to find a unique error of minimum weight that matches a given syndrome. However, our observations reveal that QAOA has the intriguing ability to identify degenerate errors of comparable weight, providing multiple potential solutions that match the given syndrome with comparable probabilities.
This is illustrated through simulations of the generator-based QAOA decoding of the [[9,1,3]] Shor code on specific error syndromes.
\end{abstract}

\keywords{degeneracy; QAOA; quantum check-based decoding; quantum generator-base decoding.}

\maketitle

\section{Introduction}

Quantum computers harness  quantum effects to perform computations beyond the capabilities of classical computers. A class of heuristic algorithms known as the quantum approximate optimization algorithm (QAOA), introduced by Farhi, Goldstone, and Gutmann \cite{FGG14a,FGG14b}, are capable of providing approximate solutions to computationally hard combinatorial optimization problems.
A combinatorial optimization problem is first embedded into a reward Hamiltonian, such that the eigenstate with the largest eigenvalue represents the optimal solution. The  Hamiltonian evolution is then approximated by executing $p$ iterations, referred to as ``levels," using two unitary operators defined by $2p$ angle parameters denoted as $({\bm\gamma,\bm\beta})$. Given these parameters determined by classical optimization, QAOA generates a distribution of potential solutions, from which one can infer an approximate solution. As the number of iterations increases, the quality of the approximate solution improves and approaches the optimal solution.
The determination of these angle parameters relies on classical computations, and thus the QAOA to typically alternate between classical and quantum computers to find a favorable solution. While employing a larger quantum circuit depth may enhance computational power \cite{CCL20,CM20}, even a single iteration of QAOA (referred to as level-1 QAOA) has the potential to demonstrate quantum advantage \cite{FH16}.

Consider the following \textit{syndrome decoding} problem.

\begin{definition}  \label{def:classical_decoding}
(Syndrome decoding of a binary linear code)
	Given a (parity-check) matrix $H\in\{0,1\}^{r\times n}$, a binary (syndrome) vector $\bmz\in\{0,1\}^{r}$, and a nonnegative integer $w$,  find a vector $  \bme\in\{0,1\}^n$ of Hamming weight no larger than $w$ such that $ {\bme}H^T =\bmz$. 
\end{definition}

\noindent 
The decision problem associated with the syndrome decoding problem  in Definition~\ref{def:classical_decoding} is one of the NP-complete problems,
which represent the most challenging NP problems,
meaning that if solutions for some NP-complete problem could be quickly found, then solutions for any other NP problem could be rapidly determined as well.
This is because both the  problems   {\it three-dimensional matching} (3D-Matching) and {\it finding the maximum cut} (Max-Cut), known as NP-complete,  can be effectively reduced to a decoding problem \cite{BMVT78,BB89}.
Since any interesting NP-hard combinatorial problem can be reduced to a syndrome decoding problem, we focus on solving this syndrome decoding problem using QAOA in this paper.

A generator matrix of a linear code is orthogonal to its parity-check matrix, providing an alternative formulation of the decoding problem in terms of the generator matrix as well \cite{MS77}.  
We will study the QAOA formulation of the decoding problem  using either a parity-check matrix or a generator matrix.

  Bruck and Blaum  have shown that an error-correcting code can be described by an energy function,  where the codewords are represented by peaks in the topography of this function \cite{BB89}.
  {A codeword-based decoding problem involves finding a vector in the code that is closest to a given corrupted codeword.} Then decoding a corrupted codeword becomes equivalent to identifying the closest peak in the energy function.
Recently, Matsumine, Koike-Akino, and Wang explored the application of QAOA to address this channel decoding problem \cite{MKAW19}. Notably, they used a reward Hamiltonian that closely resembles the energy function introduced by Bruck and Blaum \cite{BB89}.
The findings in \cite{MKAW19} suggested that reward Hamiltonians based on low-density generator matrices are particularly suitable for the level-1 QAOA. When a vector has high weight, it leads to interactions between multiple qubits in the quantum system, making a sparse matrix desirable to manage these interactions more efficiently.

When dealing with a classical block code and given a sparse generator (or parity-check) matrix, it is usually challenging to find a corresponding parity-check (or generator) matrix  that maintains the sparsity property. (This is possible for small codes as we will showcase in the numerical results.) As a result, we will create reward Hamiltonians based on the given generator or parity-check matrices for decoding.

 On the other hand, linear codes defined by low-density parity-check  (LDPC) matrices are   known as capacity-approaching codes \cite{Gal63}. 
 These LDPC codes can be efficiently decoded and are widely used in various applications.
  However, the generator matrix of an LDPC code is typically not sparse.
 As a result, a parity-check based decoder becomes necessary  for LDPC codes.
 
 In the quantum realm, binary quantum codes share similarities with classical quaternary additive codes that are dual-containing \cite{CRSS98}. We encounter a related syndrome-based (bounded-distance) decoding problem in the quantum setting, which has been proven to be NP-hard \cite{HG11,KL13_20}. Note that unless specific methods such as the Steane or Knill syndrome extraction techniques are employed \cite{Ste97L,Knill05,ZLB+20}, the general quantum decoding problem remains syndrome-based.

Moreover, quantum codes allow degenerate errors, meaning different Pauli errors may have the same effects on the code space
and they can be corrected by the same recovery operation. 
  To be more specific, a quantum stabilizer code $\cC(\cS)$ is defined by an abelian stabilizer group $\cS$ \cite{Got97}. Error operators $E$ and $E g$, for $g\in\cS$, exhibit identical  effects on the code space   and they do not need to be distinguished. (Thus they are referred to as degenerate errors to one another.)
Therefore, the optimal decoding criterion is to find  an error operator that matches a given error syndrome in a manner that maximizes the total probability of this error operator and its degenerate errors, known as the \textit{coset probability} of this error.
  The task of finding a legal error with the maximum coset probability is referred to as  degenerate syndrome decoding of a quantum stabilizer code. 
  {This problem is \#P-complete~\cite{IP15}, which is considered more difficult than NP-complete problems}.  
  Essentially, solving this problem involves enumerating the elements within an error coset.
The concept of degeneracy is not typically considered in conventional belief propagation decoders \cite{MMM04,KL20}; however, it can still be exploited in belief propagation, allowing a decoder to output degenerate errors \cite{KL21}.
 Given that QAOA can generate a probability distribution of all potential solutions, it becomes feasible to obtain each degenerate error with high probability.
Therefore, we aim to explore the application of QAOA to study and address this degeneracy decoding problem.

In this paper, we propose two reward Hamiltonians for the QAOA decoding of classical or quantum codes, based on their generator or parity-check matrices.
The generator-based Hamiltonian is similar to that in \cite{MKAW19}, following the concept introduced by Bruck and Blaum \cite{BB89}. On the other hand, the check-based Hamiltonian incorporates two essential terms: one for ensuring parity-check satisfaction and the other for managing error weight.
 We aim to favor low-weight errors that satisfy all the parity checks. To strike a balance between these two terms, we introduce two parameters that can be adjusted accordingly.
 Both Hamiltonians are more intricate when applied to quantum codes due to the consideration of generalized Hamming weight for Pauli errors.

In \cite{LP19}, Liu and Poulin  devised an objective function to train a neural network that incorporates the notion of degenerate decoding. This is achievable during the training process because the trained errors can be referenced in backward propagation. In contrast, the QAOA decoding approach commences with an equally-weighted superposition of all potential solutions. By utilizing the designed reward Hamiltonian, QAOA naturally generates degenerate errors of comparable weight, which subsequently emerge in the output distribution with comparable probabilities. As a result, QAOA inherently possesses the ability to output degenerate errors.

Finally, we conduct simulations of the level-$p$ QAOA decoding for three different codes: the classical $[7,4,3]$ Hamming code \cite{Ham50}, the unique $[[5,1,3]]$ quantum code \cite{BDSW96,LMPZ96}, and the $[[9,1,3]]$ Shor code \cite{Shor95}, with $p\leq 4$. Notably, we demonstrate that the level-$4$ check-based QAOA decoding of the $[7,4,3]$ Hamming code aligns with the optimal maximum-likelihood decoding. Considering the parity-check matrix of the $[7,4,3]$ Hamming code, which is relatively dense, we anticipate that the check-based QAOA decoding might perform well for larger codes with sparse parity-check matrices.
For the quantum case, the generator matrix of the $[[5,1,3]]$ code is dense. Nevertheless, we find that the level-$4$ generator-based QAOA decoding of the $[[5,1,3]]$ code successfully matches the optimal maximum-likelihood decoding.

 Note that defining an equivalent problem with additional redundant clauses can lead to a Hamiltonian that is easier for optimization by QAOA, while the overall complexity of the problem remains the same. We will demonstrate this through the check-based decoding of the $[7,4,3]$ code, which is a cyclic code with an $n\times n$ parity-check matrix with cyclically generated rows, providing equal protection to every bit.

We conduct simulations of the level-$p$ generator-based QAOA decoding for the $[[9,1,3]]$ Shor code, as the generator-based reward Hamiltonian requires fewer qubits compared to the check-based one. However, due to the complexity of simulating the QAOA process, it is challenging to obtain a complete decoding performance curve for $p > 2$. Instead, we compare the output distribution of QAOA to the actual conditional distribution based on the channel statistics. We demonstrate that the QAOA output distribution closely approximates the real distribution for specific error syndromes in terms of  the Jensen--Shannon divergence \cite{ES03}.

Moreover, we show that the QAOA returns $Z$ errors on any one of the first three qubits with nearly equal probabilities, as they are degenerate errors of the same weight. This finding suggests that the QAOA decoding of a quantum code is partially degenerate. Interestingly, even when $p=1$, the output solution distribution closely approximates the theoretical one, indicating the effectiveness of QAOA even with a low number of iterations.

{We remark that the problem of finding a maximum cut in a graph (MaxCut) can be reduced to a generator-based decoding problem. 
If the problem has multiple optimal solutions, it is like the degenerate decoding problem.  
We provide interesting examples that QAOA is able to find all max-cuts.	This will be discussed  in Appendix~\ref{sec:maxcut}.}

 	This paper is organized as follows. We review the basics of the QAOA in the next section.
 	In Sections~\ref{sec:QAOA_C} and~\ref{sec:QAOA_Q}, we define the generator- and check-based reward Hamiltonians for classical and quantum codes, respectively. 
Simulations of the $[7,4,3]$ Hamming code, the $[[5,1,3]]$ quantum code, and the $[[9,1,3]]$ Shor code are provided in Sec.~\ref{sec:sim}.
 Then we conclude in Section~\ref{sec:con}.

  	\section{Quantum approximate optimization algorithm}

 We consider quantum information in qubits with the computational basis $\{\ket{0},\ket{1}\}$.  
 The Pauli matrices in the computational basis are
 $ 
 {\hI}=\left[\begin{smallmatrix}
 	1 & 0 \\
 	0 & 1 
 \end{smallmatrix}\right],~ 
 \hX=\left[\begin{smallmatrix}
 	0 & 1 \\
 	1 & 0 
 \end{smallmatrix}\right],~ 
 \hat{Y}=\left[\begin{smallmatrix}
 	0 & -i \\
 	i & 0 
 \end{smallmatrix}\right],~ 
 { {\hZ}=\left[\begin{smallmatrix}
 	1 & 0 \\
 	0 & -1 
 \end{smallmatrix}\right]. }
 $ 
 The $m$-fold Pauli group $\mathcal{G}_{m}$  is
\begin{align*}
\mathcal{G}_{m} &= \big\{i^{c}M_{1}\otimes\cdots\otimes M_{m}:\, c\in \{0,1,2,3\},\, \\
                &\qquad M_{j}\in\{\hI,\hX,\hY,\hZ\} \text{ for } 1\le j\le m \big\}.
\end{align*}
 Let $ {\hX}_{j}$  ($ {\hZ_{j}}$) denote the quantum operator with ${\hX}$ (${\hZ}$) on the $j$-th qubit and identity on the others. 
 Let $I$ denote the identity operator with appropriate dimension.
 Sometimes we may omit the tensor product symbol in an $m$-fold Pauli operator. For example, $\hX\otimes \hY\otimes \hZ\otimes \hI\otimes \hI= \hX\hY\hZ \hI\hI= \hX_1 \hY_2 \hZ_3$.

 Herein a combinatorial optimization problem  with variable ${\bmx }=(x_{1},x_{2},\dots, x_{m})\in\{0,1\}^m$ 
is  as follows:
 \begin{align}
\text{maximize } C( {\bmx})=\sum_{j =1}^{q}C_{j }( {\bmx}).
 \end{align}
The objective function $C: \{0,1\}^m\rightarrow  \mathbb{R}$ is defined by $q$ clause functions  $C_j :\{0,1\}^m \rightarrow \{+1,-1\}$, where
 \begin{align}
C_{j }( {\bmx})=
 \begin{cases}
 +1,& \mbox{ if $ {\bmx}$ satisfies clause  $j $};\\
 -1,& \mbox{ otherwise.} 
 \end{cases}  \label{eq:C_J}
\end{align}

This combinatorial optimization problem  can be handled by an QAOA with $m$ qubits (as in Algorithm~\ref{Algorithm:I}).
A reward Hamiltonian   operator corresponding to the objective function $C(\bmx)$  is defined by 
\begin{align}
	\hC=\sum_{j =1}^q  {\hC}_j, \label{eq:hC}
\end{align}
where ${\hC}_j$  is a  Hermitian operator corresponding to clause  $C_j$ and is defined as
\begin{align}
 \hat{C}_j =\sum_{ { \bme}\in\{0,1\}^m} C_j ( \bme) \ket{ \bme}\bra{ \bme}, \label{eq:C_j}
\end{align}
with respect to the  computational basis vector  $\ket{\bme}$. 
We remark that our choice of reward Hamiltonian has eigenvalues $\{-q, -q+2, \dots, q-2, q\}$ because of \eq{eq:C_J},
which is different from that in \cite{FGG14a}.

Consider a Hermitian operator
 $
 \hB=\sum_{j=1}^{m} \hX_{j}.
$
The eigenvector  corresponding to the largest eigenvalue of $\hB$ is $|\psi_{0}\rangle=|+\rangle^{\otimes m}=\left(\frac{\ket{0}+\ket{1}}{\sqrt{2}}\right)^m$
and it will be the initial state to the QAOA. 

Define two unitary operators
\begin{align}
	U(\hC,\gamma)=&e^{-i\gamma \hC}=\prod_{j =1}^{q}e^{-i\gamma \hC_{j }}, \label{eq:UC} \\
U(\hB,\beta )=&e^{-i\beta \hB}=\prod_{j=1}^{m}e^{-i\beta \hX_{j}},
\end{align}
where $\gamma,\beta\in [0,\pi]$. 
Note that the energy gap between any two adjacent energy levels is two in $\hC$ or $\hB$,
so both the periods of $\gamma$ and $\beta$ are $\pi$.

 For any integer $p\geq 1$,
 a level-$p$ QAOA uses  quantum alternating operator ansatz circuits of depth $p$ to generate an angle-dependent quantum state
 \begin{align}
 |\psi_{\bm{\gamma},\bm{\beta}}\rangle=U (\hB,\beta_{p})U(\hC,\gamma_p)\cdots U(\hB,\beta_{1})U(\hC,\gamma_{1})|\psi_{0}\rangle. \label{eq:psi_ab}
 \end{align} 
 with   $2p$ angles $\bm{\gamma}=(\gamma_{1},\dots,\gamma_{p})\in [0,\pi]^p$ and $\bm{\beta}=(\beta_{1},\dots,\beta_{p})\in [0,\pi]^p$.
Then the expectation of the objective $\hC$ on this state $|\psi_{\bm{\gamma},\bm{\beta}}\rangle$ is  
 \begin{equation} \label{eq:F_p}
 F_{p}(\bm{\gamma},\bm{\beta})=\langle\psi_{\bm{\gamma},\bm{\beta}}|\hC|\psi_{\bm{\gamma},\bm{\beta}}\rangle.
 \end{equation}
                                                            
 Let   $M_{p}$ be the maximum of $F_{p}(\bm{\gamma},\bm{\beta})$ over the angles $\gamma_{1},\dots,\gamma_{p},\beta_{1},\dots,\beta_{p}$.
As $p$ increases, $M_p$ is closer to the optimal value $\max_{\bm{x}} C(\bm{x})$.   
 The procedure of a level-$p$ QAOA is given in Algorithm~\ref{Algorithm:I}.
 
 \begin{algorithm}[t]
 	\setcounter{AlgoLine}{0}
 	\Input{A reward Hamiltonian $\hC$. The number of qubits $m$.  An iteration number $T$.}
 	\Output{A distribution {over $\{0,1\}^m$}.}
 	\begin{enumerate}[1)]
  		
 		\item  (Classical computer)  
 		Determine $2p$ parameters ${\bm\gamma}=(\gamma_{1},\dots,\gamma_{p})\in[0,\pi]^p$ and ${\bm\beta}=(\beta_{1},\dots,\beta_{p})\in[0,\pi]^p$. 
 		
 		\item (Quantum computer) 
 		Construct the state $|\psi_{{\bm \gamma,\bm\beta}}\rangle$ in Eq.~(\ref{eq:psi_ab}) with input $\hC$.
 		\item (Quantum computer) 
 		 Measure $|\psi_{{\bm \gamma,\bm \beta}}\rangle$ in the computational basis and obtain outcome $\bmx\in\{0,1\}^m$.
 		\item Repeat steps {2) to 3)}  $T$ times and return the distribution of the measurement outcomes $\{\bmx\}$.

 	\end{enumerate}
 	
 	\caption{\mbox{level-$p$ QAOA with input $(\hC,m,T)$.}} \label{Algorithm:I}
 \end{algorithm}

 Finding the optimal angle sets $\{\gamma_{1},\dots,\gamma_{p}\}$ and $\{\beta_{1},\dots,\beta_{p}\}$ is a main obstacle for QAOA. One may use a fine grid method, which takes time ${O(\kappa^{2p})}$, where ${\kappa}$ is the number of possible values for each angle.
 Some optimization methods
	can also be applied and this will be discussed more in Sec.~\ref{sec:sim}.

\section{Syndrome decoding of classical linear codes} \label{sec:QAOA_C}

The goal of a classical decoding problem   is to find the most possible error, or equivalently, finding the closet codeword to a received vector. 
Herein we propose syndrome decoders by QAOA for  classical codes based on their generator or parity-check matrices.
These results will  be extended for quantum codes in the next section.

An $[n,k]$ classical binary linear code is the rowspace of a $k\times n$ generator matrix $G$
and its vectors are called \textit{codewords}.
It can  also be defined as the null space of an $r\times n$ parity-check matrix ${H}$ of rank $(n-k)$.
If a received vector $\bmy\in\{0,1\}^n$ is such that $\bmy H^T \neq \bm{0}$,
then $\bmy$ is not a codeword. For any $\bmy\in\{0,1\}^n$, the vector $\bms = \bmy H^T$ is called the \textit{error syndrome} of $\bmy$.
  Given   a syndrome vector $\bms \in\{0,1\}^{r}$, the minimum weight decoding rule is to find
\begin{align}
	\argmin_{\bme\in \{0,1\}^{n}  \text{ s.t. }  \bme H^T = \bms} \wt{\bme}, \label{eq:decoding1}
\end{align}
where the (Hamming) weight of a vector $\bmy\in\{0,1\}^n$, denoted $\wt{\bmy}$, is the number of its nonzero entries. 
The Hamming distance between two vectors $\bmx,\bmy\in\{0,1\}^n$ is $d_H(\bmx,\bmy)= \wt{\bmx-\bmy} = \wt{\bmx+\bmy}$ over binary field.

\subsection{Generator-based decoding}
\label{sec:generator_based_classical}

Suppose that  $\bmz=(z_1,\dots,z_n)\in\{0,1\}^n$ satisfies that $\bmz H^T=\bms$. 
Such $\bmz$ can be efficiently found.
Then an error vector $\bme$ with $\bme H^T = \bms$ can be written as $\bme=\bmu G+\bmz$ for some $\bmu\in\{0,1\}^k$.
Consequently, Eq.~(\ref{eq:decoding1}) is equivalent to 
\begin{align}
	\argmin_{\substack{\bmu\in \{0,1\}^{k} }} \wt{\bmu G+\bmz} &=\argmin_{\substack{\bmu\in \{0,1\}^{k} }} d_H(\bmu G+\bmz,\bm0) \notag\\
	&=\argmin_{\substack{\bmu\in \{0,1\}^{k} }} d_H(\bmu G,\bmz) \notag\\ 
	&=\argmax_{\substack{\bmu\in \{0,1\}^{k} }} \sum_{j =1}^n (1-2 z_j) (1-2[\bmu G]_j),\label{eq:decoding2}
\end{align}
where $[\bmu G]_j$ denotes the $j$-th entry of $\bmu G$ and a mapping $\{0,1\}\mapsto \{+1,-1\}$ is used in the last equality.

\begin{remark} \label{rm:z_s} 
	To find a vector  $\bmz\in\{0,1\}^n$ that matches a given syndrome $\bms\in\{0,1\}^{n-k}$, it can be done as follows. 
	Suppose that   $R$ is a  matrix  such that $H'=RH$ is in the systematic form \cite{MS77}, which is the row echelon form up to necessary column permutation such that $H' = [P, I]\in\{0,1\}^{(n-k)\times n}$.
 {Then  $[I,P^T]$ is a corresponding generator matrix.}
One can verify that $\bmz=(0,\dots,0,\bms R^T)$ is a vector that matches the syndrome.
\end{remark}

 The number of candidates $\bmu$ is $2^k$, and it would be impractical to test  them in sequence by brute force when $k$ is large. Naturally, one would like to solve the maximization problem corresponding to Eq.~(\ref{eq:decoding2}) in an equivalent energy maximization problem.
 We define a corresponding reward Hamiltonian to solve Eq.~(\ref{eq:decoding2}) by QAOA using $k$ qubits {\cite{MKAW19}}:
\begin{align}
\hat{C}=\sum_{j=1}^{n} \left( (1-2z_{j}) \prod_{\ell =1}^k \hZ_{\ell}^{[G]_{\ell,j}} \right), \label{eq:cost_generator}
\end{align}
where  $[G]_{\ell,j}$ is the $(\ell,j)$-th entry of $G$.
Note that $\hZ_{\ell}^0 =I$ and  $\hZ_{\ell}^1 = \hZ_{\ell} $.
One can verify that for $\bmu\in\{0,1\}^k$, 
\begin{align}
 \left(\prod_{\ell =1}^k \hZ_{\ell}^{[G]_{\ell,j}} \right)\ket{\bmu}= (1-  2[\bmu G]_j) \ket{\bmu}. 
\end{align}
Hence finding an eigenvector of the reward Hamiltonian $\hC$ defined in Eq.~(\ref{eq:cost_generator}) with the largest eigenvalue is equivalent to Eq.~(\ref{eq:decoding2}). 
A generator-based syndrome decoding of a classical linear code by QAOA is summarized in Algorithm~\ref{Algorithm:II}.

\begin{algorithm}[t]
	\setcounter{AlgoLine}{0}
	\Input{An $r\times n$ parity-check matrix $H$  and a syndrome $\bms\in\{0,1\}^{r}$. Iteration number $T$.}
	\Output{An error $\tilde{\bme}\in\{0,1\}^n$ such that $\tilde{\bme} H^T=\bms$.}
	\begin{enumerate}[1)]
		\item  Find a 
        generator matrix $G$ with respect to $H$ and a vector $\bmz\in\{0,1\}^n$ such that $\bmz H^T=\bms$. 
Let  $\hC$ be defined as in Eq.~(\ref{eq:cost_generator}) using $G$ and $\bmz$.
		\item Run Algorithm~\ref{Algorithm:I} with input ($\hat{C}$, $k$, $T$). 
		\item  
 Get an estimate $\tilde{\bmu}$ from the output distribution by Algorithm~\ref{Algorithm:I} and return $\tilde{\bme}=\tilde{\bmu}G+\bmz$.	
	\end{enumerate}
	\caption{Generator-based syndrome decoding of a classical code by QAOA.} \label{Algorithm:II}
\end{algorithm}

\begin{example}
	Consider the $[7,4]$ Hamming code \cite{Ham50} with generator  and parity-check matrices
 	\begin{equation}
G= \left[\begin{smallmatrix}
	1&0&0&0&1&1&0\\
	0&1&0&0&1&0&1\\
	0&0&1&0&0&1&1\\
	0&0&0&1&1&1&1
\end{smallmatrix}\right] 
~~\text{and}~~
	H=\left[\begin{smallmatrix}
	1&1&0&1&1&0&0\\
	1&0&1&1&0&1&0\\
	0&1&1&1&0&0&1\\
	\end{smallmatrix}\right].  \label{eq:H_743}
	\end{equation}
Assume that the error syndrome is $(0,1,0)$.
We find that the vector $(0,0,0,0,0,1,0)$ is of syndrome $(0,1,0)$.
Thus  the reward Hamiltonian is
\[
\hC={\hZ_{1}+\hZ_{2}+\hZ_{3}+\hZ_{4}+\hZ_{1}\hZ_{2}\hZ_{4}-\hZ_{1}\hZ_{3}\hZ_{4}+\hZ_{2}\hZ_{3}\hZ_{4}}.
\]
	
\end{example}

 \subsection{Check-based decoding}\label{sec:check_based_classical}

 We may also define an energy Hamiltonian directly based on the parity-check matrix $H$. 
 In \cite{BB89}, an energy topology is defined such that the errors corresponding to the given syndrome $\bms$ are the points in the topology with the lowest energy. However, the weight of an error is not reflected in the energy topology so a modification is necessary.

We define the following check-based reward Hamiltonian for QAOA with $n$ qubits
according to syndrome 
$\bms=(s_1 s_2 \dots s_{r})\in\{0,1\}^{r}$:
 \begin{align}
 \hC= {\eta}\sum_{j=1}^{r} (1-2s_{j})\prod_{\ell=1}^n{\hZ_{\ell}^{[H]_{j,\ell}}} + {\alpha}\sum_{j=1}^{n}{\hZ_{j}},\label{eq:cost_parity}
 \end{align}
 where the first term characterizes the parity-check satisfaction similarly to Eq.~(\ref{eq:cost_generator}) and the second term is a penalty function for error weights.
Note that $\alpha$ and $\eta$ are positive integers so that  the values of {QAOA angle parameters} $\gamma$ remain in $[0,\pi]$.

Since $\hZ=\ket{0}\bra{0}-\ket{1}\bra{1}$, every computational basis state is an eigenstate of $\sum_{j=1}^{n}{\hZ_{j}}$
and a basis vector of lower weight would have a higher energy.
 This Hamiltonian is similar to the energy topology for belief propagation decoding of quantum codes defined in \cite{KL21}.
  
  \begin{example}
  Consider the $[7,4]$ Hamming code again and assume the error syndrome $(0,1,0)$. Then  the reward Hamiltonian is
  \begin{align*}
    \hC =&  {\eta}\left(\hZ_{1}\hZ_{2}\hZ_{4}\hZ_{5}-\hZ_{1}\hZ_{3}\hZ_{4}\hZ_{6}+\hZ_{2}\hZ_{3}\hZ_{4}\hZ_{7}\right)
    +{\alpha}\sum_{j=1}^{7}\hZ_{j}.
  \end{align*} 
  \end{example}

 A check-based syndrome decoding of a classical code by QAOA is summarized in Algorithm~\ref{Algorithm:III}.
 
 \begin{algorithm}[t]
 	\setcounter{AlgoLine}{0}
 	\Input{An ${r}\times n$ parity-check matrix $H$  and a syndrome $\bms\in\{0,1\}^{r}$. Iteration number $T$. 
    Integers $\alpha,\eta >0$.}
 	\Output{An error $\tilde{\bme}\in\{0,1\}^n$ such that $\tilde{\bme} H^T=\bms$.}
 	\begin{enumerate}[1)]

 		\item  Let $\hC$ be defined as in Eq.~(\ref{eq:cost_parity}) using $H$ and $\bms$.

 		\item Run Algorithm~\ref{Algorithm:I} with input ($\hC$, $n$, $T$).
 		\item  
 		Estimate $\tilde{\bme}$ from the output distribution by Algorithm~\ref{Algorithm:I}
 		 and return $\tilde{\bme}$.		
 	\end{enumerate}
 	
 	\caption{Check-based syndrome decoding of a classical code by QAOA.} \label{Algorithm:III}
 \end{algorithm}

  	 The measurement outcomes in QAOA are error candidates and 
  	 a decision remains to be made upon receiving the distribution. A potential method is to first remove the vectors that do not match the syndrome and then choose the remaining vector with the lowest weight.

\section{Syndrome decoding of quantum stabilizer codes}\label{sec:QAOA_Q}

Suppose that $\mathcal{S}$ is an abelian subgroup of $\mathcal{G}_{n}$ such that $-{I}\notin \cS$ and let $V$ be a $2^{n}$-dimensional complex inner product space with a standard basis $\{|\bme \rangle: \bme\in\{0,1\}^n \}.$ Define 
	$\mathcal{C(S)}\equiv\{|\bmv\rangle\in V: g|\bmv\rangle=|\bmv\rangle \  \forall\, g \in \mathcal{S}\}.$
Suppose that $\cS=\langle g_{1},\dots,g_{n-k}\rangle$ has $n-k$ independent generators. Then $\mathcal{C(S)}$ is a $2^{k}$-dimensional complex vector subspace of $V$ and   is called an $[[n,k]]$ stabilizer code.

{The discretization theorem says that if  a set of error operators can be corrected by a quantum code, any linear combination of these error operators can also be corrected \cite{NC00}.
	Thus we consider quantum errors that are tensor product of Pauli matrices.}

Any two Pauli operators either commute or anticommute with each other.
A Pauli error $E\in\cG_n$ can be detected if it anticommutes with some stabilizers. So the error syndrome  of $E$ with respect to a set $\{f_{1},\dots,f_{r}\}\subset\cS$ that generate $\cS$ is defined as
 $\bms(E) = (s_{f_{1}}(E),\dots,s_{f_{r}}(E))$, where
\begin{align}
	s_{g}(E) =\begin{cases}
		0, & \text{if $gE=Eg$};\\
		1, & \text{if $gE=-Eg$}.
	\end{cases}
\end{align}

 	A Pauli operator $g\in\cG_n$ can also be represented as $g=i^c\prod_{j=1}^n \hX_j^{u_j}\hZ_j^{v_j}$, where $u_j,v_j\in\{0,1\}$ for $1\le j\le n$ and $c\in\{0,1,2,3\}$~\cite[Eq.\,(2)]{CRSS98}.
Define a homomorphism $\varphi:\mathcal{G}_{n}\mapsto \mathbb{Z}_{2}^{2n}$ on $g$ by $\varphi(g)=(u_{i},\dots,u_{n}|v_{1},\dots,v_{n})$.
When the phase of a Pauli operator is irrelevant, it suffices to discuss its corresponding binary $2n$-tuple.
Thus a check matrix corresponding to  $\{f_{1},\dots,f_{r}\}$  is defined  as
\begin{align}
	{H}_{\cS}\equiv\left[\begin{smallmatrix}
		\varphi(f_{1})\\
		\varphi(f_{2})\\
		\vdots\\
		\varphi(f_{r})\\
	\end{smallmatrix}\right], \label{eq:check_matrix}
\end{align} which is an $r\times 2n$ binary matrix. 
Consequently, the error syndrome of a Pauli operator $E\in\cG_n$ is 
\begin{align}
	\bms(E) = \varphi(E)\Lambda {H}_{\cS}^{T}, \label{eq:symp_inn_prod}
\end{align}
where  $\Lambda=\left[\begin{smallmatrix}
	O_{n\times n } & {I}_{n\times n}\\
	{I}_{n\times n} & O_{n\times n }
\end{smallmatrix}\right]$, $I_{n\times n}$ is the $n\times n$ identity matrix, and $O_{n\times n }$ is an $n\times n$ all-zero matrix. 
Therefore, one can equivalently consider a binary code of length $2n$.
However, the notion of weight is different; the weight of a Pauli operator $E$ is  the \textit{generalized weight} $\gw{\varphi(E)}$ of  the binary $2n$-tuple $\varphi(E)$, defined by  
\begin{align}
	\gw{\bmv}\equiv  \left|\{j\in\{1,\dots,n\}: v_j=1 \mbox{ or } v_{n+j}=1 \}\right|	\label{eq:gw}
\end{align}
for $\bmv=(v_1,v_2,\dots, v_{2n})\in\{0,1\}^{2n}$.
When $n=1$, we have $\gw{0|0}=0$ and $\gw{0|1}=\gw{1|0}=\gw{1|1}=1$.

The normalizer group of $\cS$ in $\cG_n$ is
$$
N(\cS)=\{ g\in\cG_n: gh=hg, \, \text{ for all } h\in\cS\}.
$$
{Note that $N(\cS)$ is also the centralizer group (or called commutator group) of $\cS$ in $\cG_n$. 
This is because two elements in $\cG_n$ either commute or anticommute and $-I\notin \cS$.}
One can find  $g_{n-k+1},\dots,g_n,$ $h_{n-k+1},\dots, h_n\in\cG_n$ such that 
$N(\cS)$ is generated by $\{g_1,\dots,g_n, h_{n-k+1},\dots,h_n\}$ up to phases \cite{CG97,NC00}.
For $E\in\cG_n$, one readily understand that $gE$ and $E$ have the same error syndrome for any $g\in N(\cS)$.
Thus, there are $2^{n+k}$ errors, up to phases, with the same syndrome.
Moreover, for $g\in\cS$,  $gE$ and $E$ have the same effects on the codespace and thus they are called \textit{degenerate errors} of each other.
The minimum distance of a quantum code is 
	$$
	d = \min\{\gw{\varphi(g)}:\, g\in N(\cS),~ g\notin\{\pm 1,\pm i\}\times\cS\}.
	$$
An $[[n,k,d]]$ quantum code is called a degenerate code if there exists nonidentity $g\in\cS$ with $\gw{\varphi(g)}< d$.

 \subsection{Generator-based decoding}
 
  	We  will similarly devise a  generator-based syndrome decoding of a quantum stabilizer code by QAOA.
 This is done by considering  the dual of the  binary code of length $2n$ defined by the check matrix~\eq{eq:check_matrix} of an $n$-qubit quantum code.

Suppose that   
$N(\cS)$ is generated by $\{g_1,\dots,g_n,$ $ h_{n-k+1},\dots,h_n\}$ up to phases.
Consider an $(n+k)\times 2n$ binary matrix 
	\begin{equation} \label{eq:G_S}
	G_{\cS}=\left[\begin{smallmatrix}
		\varphi( g_{1})\\
		\vdots\\
		\varphi(g_{n})\\
		\varphi(h_{n-k+1})\\
		\vdots\\
		\varphi(h_{n})\\
	\end{smallmatrix}\right],
	\end{equation}
which satisfies $G_{\cS}\Lambda H_{\cS}^T = O$.
Such $G_\cS$ can be efficiently found, for example, by {the standard form} of the stabilizer code \cite{CG97,NC00,KL19}.
Given a syndrome $\bms=(s_1,\dots,s_r)\in\{0,1\}^{r}$, one can efficiently find $\bmz=(z_1,\dots,z_{2n})\in\{0,1\}^{2n}$ such that 
$\bmz\Lambda H_{\cS}^T=\bms.$
Consequently a potential error is of the form 
\[
\bmz+ \bmu G_{\cS}
\]
for $\bmu\in\{0,1\}^{n+k}$.
We can similarly define a reward Hamiltonian as in Sec.~\ref{sec:generator_based_classical}.
However, the notion of weight is different and we have to design a Hamiltonian regarding to the generalized weight.
We define a generalized distance 
	$d_{\text{gw}}(\bma,\bmb)= \gw{\bma+\bmb}$ for $\bma,\bmb\in\{0,1\}^{2n}$.
Then the minimum weight decision rule is 
\begin{align}
	&\argmin_{\substack{\bmu\in \{0,1\}^{n+k} }} \gw{\bmu G_\cS+\bmz}\notag\\
	=&\argmin_{\substack{\bmu\in \{0,1\}^{n+k} }} d_{\text{gw}} (\bmu G_\cS,\bmz) \notag \\
	=&\argmin_{\substack{\bmu\in \{0,1\}^{n+k} }} \sum_{j =1}^n  d_{\text{gw}} (([\bmu G_\cS]_j,[\bmu G_\cS]_{n+j}), (z_j,z_{n+j})) \notag \\
	=&\argmax_{\substack{\bmu\in \{0,1\}^{n+k} }} \sum_{j =1}^n  (-1)^{d_{\text{gw}} (([\bmu G_\cS]_j,[\bmu G_\cS]_{n+j}), (z_j,z_{n+j}))},
\label{eq:decoding31}
\end{align}
where   a mapping $\{0,1\}\mapsto \{+1,-1\}$ is used in the last equality. 
For $a_1,a_2,b_1,b_2\in\{0,1\}$, $d_\gw {(a_1,a_2), (b_1,b_2)}= 1$ if $a_j\neq b_j$ for some $j\in\{1,2\}$, 
and $d_\gw{(a_1,a_2), (b_1,b_2)}= 0$, otherwise.
One can verify that 
\begin{align}
&(-1)^{d_\gw{ (a_1,a_2), (b_1,b_2) }} \notag \\
=& \frac{1}{2}\left( (-1)^{a_1+b_1}+(-1)^{a_2+b_2}+ (-1)^{a_1+b_1+a_2+b_2} -1   \right).
\end{align}
Thus Eq.~(\ref{eq:decoding31}) can be rewritten as
\small
\begin{align}
 \argmax_{\substack{\bmu\in \{0,1\}^{n+k} }} & \frac{1}{2}\sum_{j =1}^n \left((-1)^{z_j}(-1)^{[\bmu G_\cS]_j} + (-1)^{z_{n+j}}(-1)^{[\bmu G_\cS]_{n+j}}\right. \notag \\
                                                                        &\left.+(-1)^{z_j}(-1)^{z_{n+j}}(-1)^{[\bmu G_\cS]_j}(-1)^{[\bmu G_\cS]_{n+j}} -1\right).
	\label{eq:decoding3}
\end{align}
\normalsize

Thus we define a generator-based reward Hamiltonian for a quantum stabilizer code with syndrome $\bms$ 
as 
\small
\begin{align}
	\hat{C}=& \frac{1}{2}\sum_{j=1}^{n}\left((1-2z_{j})\prod_{\ell =1}^{n+k} \hZ_{\ell}^{[G_\cS]_{\ell,j}}
 +  (1- 2z_{n+j})\prod_{\ell =1}^{n+k} \hZ_{\ell}^{[G_\cS]_{\ell,n+j}} \right.\notag \\&
	+\left.(1-2z_{j})(1- 2z_{n+j})\prod_{\ell =1}^{n+k} \hZ_{\ell}^{[G_\cS]_{\ell,j}}\prod_{\ell =1}^{n+k} \hZ_{\ell}^{[G_\cS]_{\ell,n+j}}-I \right), \label{eq:cost_generator_quantum}
\end{align}
\normalsize
which is defined according to Eq.~(\ref{eq:decoding3}).

A generator-based syndrome decoding of a quantum stabilizer code by QAOA is summarized in Algorithm~\ref{Algorithm:IV}.
\begin{algorithm}[t]
	\setcounter{AlgoLine}{0}
	\Input{An $r\times 2n$ check matrix $H_{\cS}$  and a syndrome $\bms\in\{0,1\}^{r}$. Iteration number~$T$.}
	\Output{An error $\tilde{\bme}\in\{0,1\}^{2n}$ such that $\tilde{\bme} {\Lambda} H_\cS^T=\bms$.}
	\begin{enumerate}[1)]
		\item  Find a generator matrix $G_\cS$ with respect to $H_\cS$ and a vector $\bmz\in\{0,1\}^{2n}$ such that $\bmz {\Lambda} H_{\cS}^T=\bms$. 
		Let  $\hC$ be defined as in Eq.~(\ref{eq:cost_generator_quantum}) using $G_{\cS}$ and $\bmz$.
		\item Run Algorithm~\ref{Algorithm:I} with input ($\hat{C}$, {$n+k$}, $T$).
		\item  
		Get an estimate $\tilde{\bmu}$ from the output distribution by  Algorithm~\ref{Algorithm:I} and return $\tilde{\bme}=\tilde{\bmu}G_{\cS}+\bmz$.
	\end{enumerate}

	\caption{Generator-based syndrome decoding of a quantum stabilizer code by QAOA.} \label{Algorithm:IV}
\end{algorithm}

 \begin{example} \label{ex:513}
 The  $[[5,1,3]]$ code \cite{BDSW96,LMPZ96} is defined by the stabilizer group $\mathcal{S}=\langle {\hX\hZ\hZ\hX\hI},{\hI\hX\hZ\hZ\hX},$ ${\hX\hI\hX\hZ\hZ},$ ${\hZ\hX\hI\hX\hZ}\rangle\subset\cG_5 $ 
 and $N(\cS)=\{\pm 1,\pm i\} \times \langle {\hX\hZ\hZ\hX\hI},$ ${\hI\hX\hZ\hZ\hX},$ ${\hX\hI\hX\hZ\hZ},$ ${\hZ\hX\hI\hX\hZ},$ $  \hZ^{\otimes 5},$ $\hX^{\otimes 5} \rangle$~\cite{Got97}. 
 Thus
 \begin{align*}
 	{H}_{\cS}= \left[\begin{smallmatrix}
 		1 & 0 & 0 & 1 & 0\\
 		0 & 1 & 0 & 0 & 1\\
 		1 & 0 & 1 & 0 & 0\\
 		0 & 1 & 0 & 1 & 0
 	\end{smallmatrix}\middle| \begin{smallmatrix}
 		0 & 1 & 1 & 0 & 0\\
 		0 & 0 & 1 & 1 & 0\\
 		0 & 0 & 0 & 1 & 1\\
 		1 & 0 & 0 & 0 & 1 
 	\end{smallmatrix}\right]
    ~\text{and}~
	{G}_{\cS}= \left[\begin{smallmatrix}
 		&&&&&H_\cS&&&&\\
		\hline
		0 & 0 & 0 & 0 & 0  & ~~1 & 1 & 1 & 1 & 1 \\
		1 & 1 & 1 & 1 & 1  & ~~0 & 0 & 0 & 0 & 0
	\end{smallmatrix}\right].
 \end{align*}
Suppose that the syndrome is $(0,0,0,1)$. One can check that $(1,0,0,0,0\,|\,0,0,0,0,0)$ matches this syndrome. The reward Hamiltonian is 
\begin{align*} 
	\hC= \frac{1}{2}\Big(
		&(-\hZ_{1}\hZ_{3}\hZ_{6}+\hZ_{4}\hZ_{5}-\hZ_{1}\hZ_{3}\hZ_{4}\hZ_{5}\hZ_{6}-I)\\
		& +(\hZ_{2}\hZ_{4}\hZ_{6}+\hZ_{1}\hZ_{5}+\hZ_{1}\hZ_{2}\hZ_{4}\hZ_{5}\hZ_{6}-I)\\
		&+(\hZ_{3}\hZ_{6}+\hZ_1\hZ_{2}\hZ_{5}+\hZ_{1}\hZ_{2}\hZ_{3}\hZ_{5}\hZ_{6}-I)\\
		& +(\hZ_{1}\hZ_{4}\hZ_{6}+\hZ_2\hZ_{3}\hZ_{5}+\hZ_{1}\hZ_{2}\hZ_{3}\hZ_{4}\hZ_{5}\hZ_{6}-I)\\
		&+(\hZ_{2}\hZ_{6}+\hZ_{3}\hZ_{4}\hZ_{5}+\hZ_{2}\hZ_{3}\hZ_{4}\hZ_{5}\hZ_{6}-I)
		\Big).
\end{align*}
 \end{example}

 	  \subsection{Check-based decoding}

 Next,	we  would like to devise a check-based syndrome decoding of  a quantum stabilizer code by QAOA.
 This is done by analyzing the binary code of length $2n$ defined by the check matrix~\eq{eq:check_matrix} of an $n$-qubit quantum code.
 We have to design a penalty term regarding the generalized weight of a binary $2n$-tuple. 
 
 Consider the following operator
\begin{align}
\hat{W}=&\frac{\hat{Z}_1+\hZ_2+\hZ_1\hZ_2-I}{2}\notag \\
=& \ket{00}\bra{00}-\ket{01}\bra{01}-\ket{10}\bra{10}-\ket{11}\bra{11}.
\end{align}
One can see that $01,$ $10$, $11$ are of generalized weight one and their corresponding basis states are the eigenstates of $\hat{W}$ with eigenvalue $-1$.
Therefore,  we define a check-based reward Hamiltonian according to syndrome $\bms=(s_1,\dots,s_{r})\in\{0,1\}^{r}$  and Eq.~(\ref{eq:check_matrix}) as
 	\begin{align}\hC=& {\eta}\sum_{j=1}^{r}(1-2s_{j})\prod_{\ell=1}^{2n}{\hZ_{\ell}^{[H_{\cS}\Lambda]_{j,\ell}}}\notag \\
 		&+\frac{\alpha}{2}\sum_{j=1}^{n}\left({\hZ_j+\hZ_{n+j}+\hZ_{j}\hZ_{n+j}-I}\right). \label{eq:cost_parity_quantum}	
 	\end{align} 
 Note that $\alpha$ and $\eta$ are positive integers so that  the values of {QAOA parameters} $\gamma$ remain in $[0,\pi]$.  

\begin{example}
	  Consider the $[[5,1,3]]$  code with syndrome $(0,0,0,1)$  and
	$${H}_{\cS}\Lambda= \left[\begin{smallmatrix}
		0 & 1 & 1 & 0 & 0\\
		0 & 0 & 1 & 1 & 0\\
		0 & 0 & 0 & 1 & 1\\
		1 & 0 & 0 & 0 & 1
	\end{smallmatrix}\middle| \begin{smallmatrix}
		1 & 0 & 0 & 1 & 0\\
		0 & 1 & 0 & 0 & 1\\
		1 & 0 & 1 & 0 & 0\\
		0 & 1 & 0 & 1 & 0 
	\end{smallmatrix}\right].$$
Thus  the reward Hamiltonian is
\small
\begin{align*}
\hC=&{\eta}\left({\hZ_{2}}{\hZ_{3}}\hZ_{6}{\hZ_{9}}+{\hZ_{3}}{\hZ_{4}}{\hZ_{7}}\hZ_{10}+{\hZ_{4}}{\hZ_{5}}{\hZ_{6}}{\hZ_{8}}-{\hZ_{1}}{\hZ_{5}}{\hZ_{7}}{\hZ_{9}}\right)\\
&+\frac{\alpha}{2}  \sum_{j=1}^{5}\left(\hZ_j+\hZ_{5+j}+\hZ_{j}\hZ_{5+j}-I\right).
\end{align*}
\normalsize
	\end{example}

A check-based syndrome decoding of a quantum stabilizer code by QAOA is summarized in Algorithm~\ref{Algorithm:V}.
 \begin{algorithm}[t]
 	\setcounter{AlgoLine}{0}
 	\Input{An ${r}\times 2n$ check matrix $H_{\cS}$  and a syndrome $\bms\in\{0,1\}^{r}$. Iteration number $T$. {Integers $\alpha,\eta>0$}. }
 	\Output{An error $\tilde{\bme}\in\{0,1\}^{2n}$ such that $\tilde{\bme} \Lambda H_\cS^T=\bms$.}
 	\begin{enumerate}[1)]
 		\item  Let $\hC$ be defined as in Eq.~(\ref{eq:cost_parity_quantum}) using $H_\cS$ and $\bms$.
 		\item Run Algorithm~\ref{Algorithm:I} with input ($\hC$, $2n$, $T$).
 		\item  
 		Estimate $\tilde{\bme}$ from the output distribution by Algorithm~\ref{Algorithm:I} 
 		and return $\tilde{\bme}$.		

 	\end{enumerate}
 	
 	\caption{Check-based syndrome decoding of a quantum stabilizer code by QAOA.} \label{Algorithm:V}
 \end{algorithm}

\section{Numerical Simulations} \label{sec:sim}

We simulate  the syndrome decoding of QAOA using classical computers.
Only small codes are considered here due to the simulation complexity that is exponential in the number of qubits.
{The performance of QAOA decoder relies on the sparsity of the generator or parity-check matrices and this will be discussed.}

A level-$p$ QAOA has  $2p$ parameters $({\bm \gamma},{\bm \beta})\in[0,\pi]^{2p}$ that need to be determined by a classical computer, and derivative-free optimization methods are preferred \cite{SSL19}.
\rmark{We will consider and compare two derivative-free methods: {\it Nelder-Mead} (NM) method \cite{NM65} and {\it constrained optimization by linear approximation} (COBYLA) \cite{Pow78,Pow94}.}

\rmark{The NM method is an optimization algorithm for finding the minimum (or maximum) of an objective function in multidimensional space. It starts by constructing a simplex, which is a geometrical figure with $n+1$ vertices in $n$-dimensional space. 
The algorithm iteratively updates the simplex vertices through certain operations until the vertices converge, indicating the minimum of the objective function.

COBYLA is an optimization algorithm designed for solving nonlinear optimization problems with constraints. It is particularly useful when the objective function and constraints are not differentiable, or when gradient information is not available.

Both NM and COBYLA do not guarantee finding a global optimum. To improve their performance, we combine them with heuristic methods such as the multistart method \cite{SSL19} or basin-hopping \cite{WD97}.
The multistart method functions like an $\varepsilon$-net, dividing the search space into multiple grids and assigning a starting point in each grid for searching.
Basin-hopping is a global optimization technique that integrates local optimization with random searches to find the global optimum of an objective function. It is particularly effective for rugged landscapes with multiple local optima. The process starts with local optimization from an initial point to find a local minimum, followed by random perturbations to the solution and subsequent local optimizations to identify new local minima.
}


\bmark{
We use the software {\tt SciPy} \cite{Vir+20} for simulations, which  supports NM and COBYLA.
We conduct a fine-grid search with $\kappa^{2p}$ multistart points, where $\kappa$ is the number of cuts for each dimension of the search space.  
	We observe that   this approach is usually sufficient to find a set of nearly optimal parameters with {small} values of $\kappa=2$ or $3$. However, the number of search points {$\kappa^{2p}$} still grows exponentially with the level $p$. Therefore, we may alternatively run basin-hopping, which is also supported in {\tt SciPy}. The number of maximum hopping iterations is fixed at 100,  regardless of the value of $p$.
}

In our simulations, we consider scenarios up to $p=4$, leading to $\kappa^{2p}=256$ if $\kappa=2$ or $\kappa^{2p}=6561$ if $\kappa=3$ for the multistart method. To maintain a comparable level of complexity with basin-hopping, we enforce a constraint that $\kappa^{2p}\le 256$ in the multistart case.

Additionally, we will normalize the maximum value of $F_p$ to one in the main text for ease of observation.

 The performance with basin-hopping  is comparable to the multistart case in our simulations.

From numerous simulations, we conclude that optimization over the angles $\bm \beta$ and $\bm \gamma$ by  NM+basin-hopping or COBYLA+multistart have better numerical results.
We will focus on the two combinations in the following.

    \subsection{Decoding classical codes by QAOA}

\begin{figure*}
	\begin{subfigure}[b]{0.245\textwidth}
		\centering
		\!\!\!\!\!\includegraphics[width=1.14\textwidth]{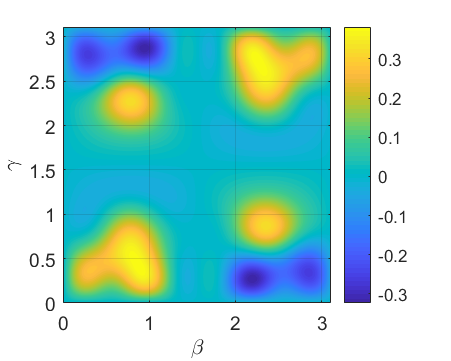}
		\caption{$(\alpha,\eta)=(1,1)$.}  \label{fig:H743_a1e1}
	\end{subfigure}
	\begin{subfigure}[b]{0.245\textwidth}
		\centering
		\!\!\!\includegraphics[width=1.14\textwidth]{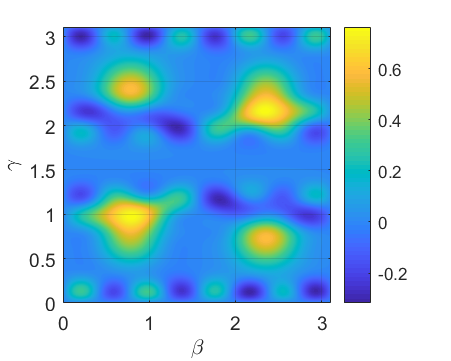}
		\caption{$(\alpha,\eta)=(1,3)$.}  \label{fig:H743_a1e3}
	\end{subfigure}
	\begin{subfigure}[b]{0.245\textwidth}
		\centering
		\!\!~\includegraphics[width=1.14\textwidth]{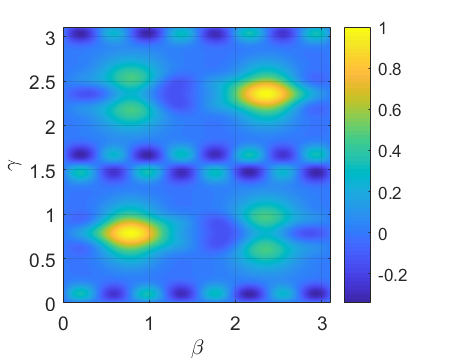}
		\caption{$(\alpha,\eta)=(1,4)$.}  \label{fig:H743_a1e4}
	\end{subfigure}
	\begin{subfigure}[b]{0.245\textwidth}
		\centering
		\!\!~~\includegraphics[width=1.14\textwidth]{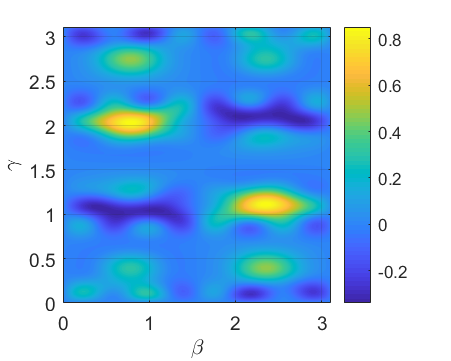}
		\caption{$(\alpha,\eta)=(2,3)$.}  \label{fig:H743_a2e3}
	\end{subfigure}
	\caption{
 Normalized $F_1$ for the reward Hamiltonian defined by $H_{[7,4,3]}$ with syndrome zero for various combinations of $(\alpha,\eta)$.
 \rmark{For each subfigure, the color on the square represents the level of $F_1$ for a given $(\gamma,\beta)\in[0,\pi]^2$, with the value of the level  indicated by a bar next to the square. The range of the levels in each bar may differ. Only the case $(\alpha,\eta)=(1,4)$ achieves a normalized $F_1$ value of 1. This case also contains many local optima with $F_p>0.5$, which is  better than the a global optimum $F_p<0.4$ in the case $(\alpha,\eta)=(1,1)$.}
	}	\label{fig:H743_aXeY}
\end{figure*}

We have both generator- and check-based QAOA decoders.
It has been demonstrated in \cite{MKAW19} that the reward Hamiltonian induced from a sparse matrix would be more suitable for QAOA.      
 Herein, we demonstrate the check-based syndrome decoding by  Algorithm~\ref{Algorithm:III},
 along with the discussion on the two parameters $\alpha$ and $\eta$.
 The simulation of the generator-based decoding  is similar  and even simpler.
 
We study the classical $[7,4,3]$ Hamming code  with a parity-check matrix 
\begin{equation} \label{eq:H_[743]}
	{H_{[7,4,3]}}=\left[\begin{smallmatrix}
		1&1&0&1&1&0&0\\
		1&0&1&1&0&1&0\\
		0&1&1&1&0&0&1\\
	\end{smallmatrix}\right]
\end{equation} 
 as given in \eq{eq:H_743}, which is not sparse since there are more ones than zeros. 
If the QAOA decoder based on this parity-check matrix   performs well,  then one would expect this QAOA decoder to work for much larger sparse matrices.

Observe from \eq{eq:cost_parity} that the eigenvalues of the two terms in the reward Hamiltonian 
are $\eta r, \eta(r-2),$ $\eta(r-4), \dots, -\eta r$ 
and $\alpha n, \alpha(n-2), \alpha(n-4), \dots, -\alpha n$
for $(n,r)=(7,3)$. 
If the syndrome is $\bm{0}$, the all-zero error vector has the highest eigenvalue $\eta r+ \alpha n$ and is preferred as we desire.
If the syndrome is nonzero, a weight-one error matching the syndrome is preferred than the all-zero vector. So we must have 
\begin{align}
	&r\eta+(n-2)\alpha \ge (r-2)\eta+n\alpha, \quad\text{i.e.,}  
&\eta \ge \alpha.	\label{eq:alpha-eta_rule}
\end{align}
On the other hand, $\eta/\alpha$ cannot be too large; otherwise, the preference for a low-eight error is diminished. 

 For a level-1 QAOA with syndrome zero in Algorithm~\ref{Algorithm:III}, the expectation values of the objective $F_1$,  normalized by $r\eta  + n\alpha $ for various  combinations of $(\alpha,\eta)$, are shown in Fig.~\ref{fig:H743_aXeY}.
As can be seen that  the normalized $F_1$ corresponding to $(\alpha,\eta)=(1,4)$ has maximum equal to one.

	\subsubsection{Using a full-rank parity-check matrix}

Consider the reward Hamiltonian defined by~$H_{[7,4,3]}$, $\alpha=1,\,\eta=4$, and some $\bms\in\{0,1\}^3$ and sufficiently large~$T$.

\begin{figure}
		\centering \includegraphics[width=0.5\textwidth]{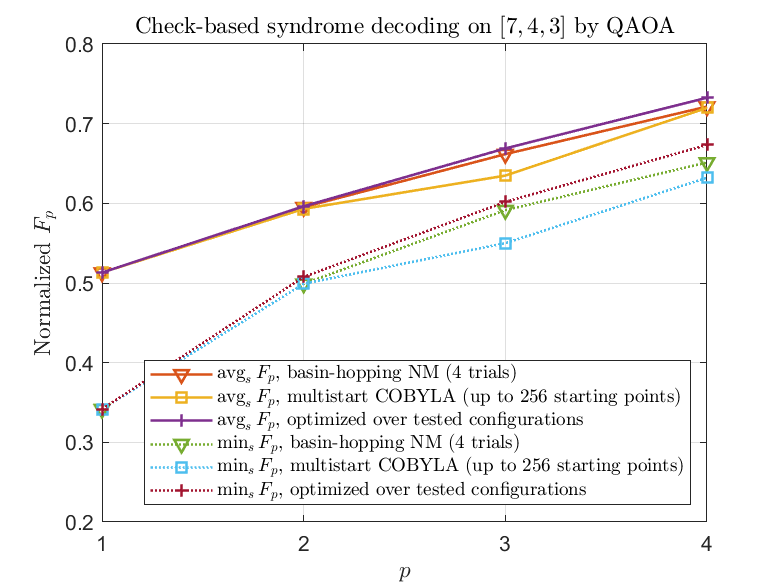}
		\caption{Simulated $F_p$ for check-based syndrome decoding of the $[7,4,3]$ code by QAOA.} \label{fig:743_Fp}		
	\end{figure}

For NM+basin-hopping, we additionally test four starting points with $(\gamma_p,\beta_p)=(0,0)$, $(\frac{\pi}{8},\frac{\pi}{8})$, $(1,1)$ or at random, and choose the one that has the largest mean objective $F_p$.
For COBYLA+multistart, each angle $\gamma_j$ or $\beta_j$ has possible $\kappa$ values, so a total of $\kappa^{2p}$ starting points
are tested. However, we choose different values of $\kappa$ such that $\kappa^{2p}\le 256$.
We do these optimizations for each syndrome $\bms$.
The results are shown in Fig.~\ref{fig:743_Fp}, where $\avg_\bms F_p$ denotes the mean expectation value $F_p$ averaged over all $\bms\in\{0,1\}^3$ and {$\min_\bms F_p$ denotes  the worst result of the tested {syndromes}.}
(The value $\min_\bms F_p$ is the expectation value corresponding to the reward Hamiltonian of some error syndrome  that is the most difficult for the QAOA.)
One can see that NM+basin-hopping performs slightly better than COBYLA+multistart.

The complete QAOA decoding of the $[7,4,3]$ code is as follows. 
An  error $\bme\in\{0,1\}^n$ is generated according to an independent binary symmetric channel (BSC) with   cross error rate $\epsilon$.
If the syndrome of the error is  zero, the decoding output  is $\hat\bme=\bm{0}$.
 Otherwise, we simulate the performance of Algorithm~\ref{Algorithm:III} with better $(\bm\gamma,\bm\beta)$ determined above.
Upon receiving an output distribution together with $T$ possible errors for a given syndrome,
the error of minimum weight that matches the error syndrome 
will be returned as the output $\hat{\bme}$. 
If there are no errors matching the syndrome, the decoding output is chosen to be $\hat\bme=\bm{0}$.
A decoding error occurs if $\hat\bme \ne \bme$.

\begin{figure}
	\centering \includegraphics[width=0.5\textwidth]{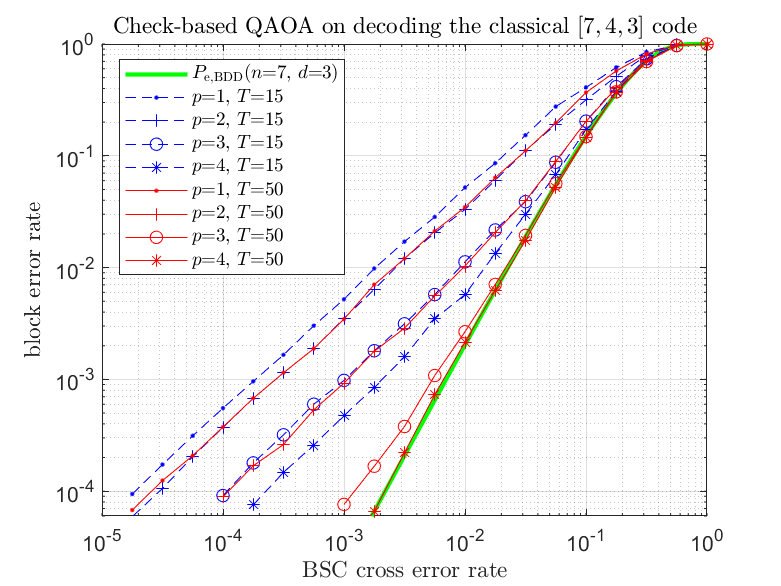}
	\caption{
		The performance of check-based syndrome decoding on the $[7,4,3]$ code by QAOA using $H_{[7,4,3]}$.
} \label{fig:743_dec}		
\end{figure}

In our simulations of logical error rate, unless otherwise specified, Monte Carlo test is performed and 500 decoding errors are collected for each data point.
The performance of QAOA decoding on the $[7,4,3]$ code is shown in Fig.~\ref{fig:743_dec} for several levels $p$ and iteration numbers $T$.
Since the $[7,4,3]$ code is perfect, the maximum-likelihood decoding rule is equivalent to  the bounded-distance decoder (BDD)
with block error rate  
	\begin{equation} \label{eq:P_e,BDD}
	P_{\rm e,BDD}(n,d) = 1- \textstyle \sum_{~j=0}^{\lfloor\frac{d-1}{2}\rfloor} \binom{n}{j} \epsilon^j (1-\epsilon)^{n-j}
	\end{equation}
at cross error rate $\epsilon$. This is also plotted in Fig.~\ref{fig:743_dec}.

For $T=15$, the block error rate is less than the cross error rate when $p\ge 4$
so that  the advantage of coding can be observed. For $T=50$, the QAOA decoder matches the performance of the BDD when $p\geq 4$.

\subsubsection{Improvement with redundant clauses}

In this subsection, we show that  QAOA decoding can be improved with a modified reward Hamiltonian that defines an equivalent decoding problem without any cost.
More precisely, we may introduce redundant  clauses but the number of qubits remains the same.
Note that this technique can be applied to either the check-based  or the generator-based decoding.
We demonstrate this on the decoding of the $[7,4,3]$ code again.

\begin{figure}
	\centering
	\begin{subfigure}[b]{0.235\textwidth}
		\centering
		\!\!\!\!\!\includegraphics[width=1.14\textwidth]{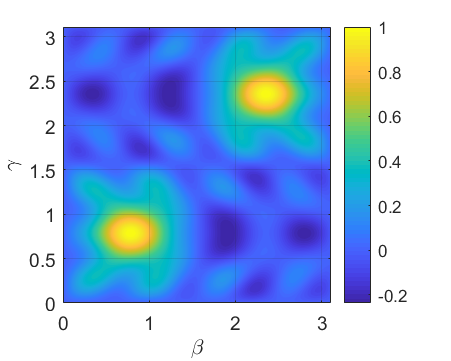}
		\caption{$(\alpha,\eta)=(1,1)$.}  \label{fig:Hc_743_a1e1}
	\end{subfigure}
	\begin{subfigure}[b]{0.235\textwidth}
		\centering
		\!\!\!\includegraphics[width=1.14\textwidth]{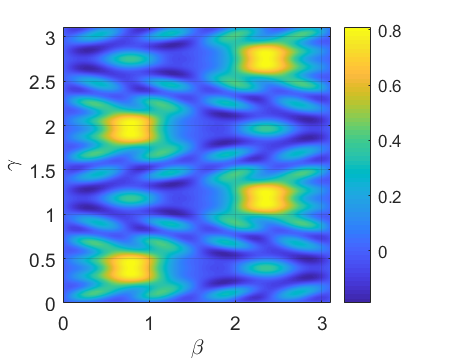}
		\caption{$(\alpha,\eta)=(2,3)$.}  \label{fig:Hc_743_a2e3}
	\end{subfigure}
	\caption{ Normalized  expectation $F_1$  for the reward Hamiltonian defined by $H^\text{(circ)}$ and $(\alpha,\eta)=(1,1)$ or $(2,3)$   with syndrome zero.}	\label{fig:Hc_743_aXeY}
    \vspace*{\floatsep}
\end{figure}

The $[7,4,3]$ code  has an equivalent code with a parity-check matrix that can be cyclicly generated by a row vector $(1,0,1,1,1,0,0)$ as
\begin{equation} \label{eq:H^c}
	H^\text{(circ)}=\left[\begin{smallmatrix}
		1&0&1&1&1&0&0\\
		0&1&0&1&1&1&0\\
		0&0&1&0&1&1&1\\
		1&0&0&1&0&1&1\\
		1&1&0&0&1&0&1\\
		1&1&1&0&0&1&0\\
		0&1&1&1&0&0&1
	\end{smallmatrix}\right].
\end{equation}
Note that $H^\text{(circ)}$ is of rank $3$. 
It is evident that $H^\text{(circ)}$ offers equal protection capability to each bit, resulting in a reward Hamiltonian with a circulant symmetry when defined according to $H^\text{(circ)}$. 
Using a similar argument as in \eq{eq:alpha-eta_rule}, we consider two cases for $(\alpha,\eta)$: $(1,1)$ or $(2,3)$. 
After examining Fig.~\ref{fig:Hc_743_aXeY}, we choose $(\alpha,\eta)=(1,1)$ due to its smoother energy topology and a maximum normalized value of one.

\begin{figure}
	\centering \includegraphics[width=0.5\textwidth]{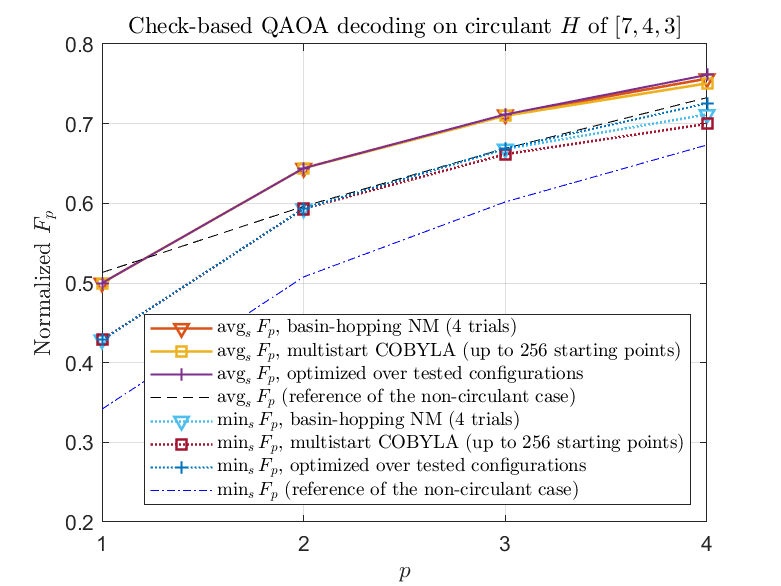}
	\caption{
		Simulated $F_p$ for check-based syndrome decoding on the $[7,4,3]$ code using the circulant $H^{\text{(circ)}}$
		or the non-circulant $H_{[7,4,3]}$. 
	} \label{fig:743_c_Fp}		\vspace*{\floatsep}
	\centering \includegraphics[width=0.5\textwidth]{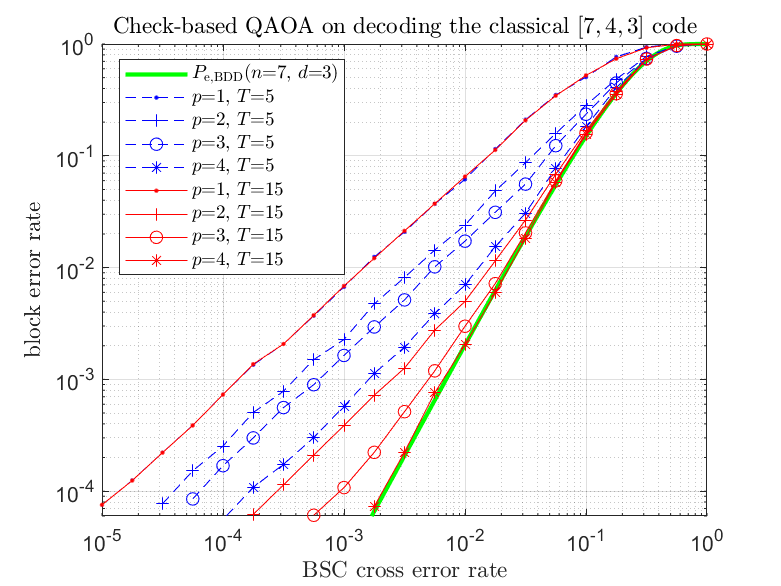}
	\caption{
		The performance of check-based syndrome decoding on the $[7,4,3]$ code by QAOA using $H^\text{(circ)}$.
	} \label{fig:743_c_dec}		
\end{figure}

	As in the previous subsection, we use NM+basin-hopping and COBYLA+multistart to optimize $F_p$ over  $(\bm\gamma,\bm\beta)\in[0,\pi]^{2p}$ for $p=1,\dots,4$ for different syndromes and the normalized $F_p$ are shown in Fig.~\ref{fig:743_c_Fp}.
	We plot also the optimized curves in Fig.~\ref{fig:743_Fp} by the non-circulant matrix $H_{[7,4,3]}$  for comparison. 
The performance is significantly improved such that  $\min_{\bms} F_p$ by $H^\text{(circ)}$ 
is comparable to $\text{avg}_{\bms}\, F_p$ by $H_{[7,4,3]}$.

The decoding performance of Algorithm~\ref{Algorithm:III} based on  $H^\text{(circ)}$ and optimized $(\bm\gamma,\bm\beta)$  is shown in Fig.~\ref{fig:743_c_dec}.
The decoding gain appears for  $T=5$
and the BDD performance is achieved for $T=15$.
These significantly improve the results in Fig.~\ref{fig:743_dec}.

\subsection{Decoding quantum codes by QAOA} \label{sec:decQ}

In this section, we simulate QAOA decoding on the $[[5,1,3]]$ and $[[9,1,3]]$ quantum codes.
Algorithm~\ref{Algorithm:IV} needs $n+k$ qubits to decode an $[[n,k]]$ code,
while Algorithm~\ref{Algorithm:V} needs $2n$ qubits.
We simulate only Algorithm~\ref{Algorithm:IV} in this section.

\subsubsection{The $[[5,1,3]]$ code}
\begin{figure}
	\centering \includegraphics[width=0.5\textwidth]{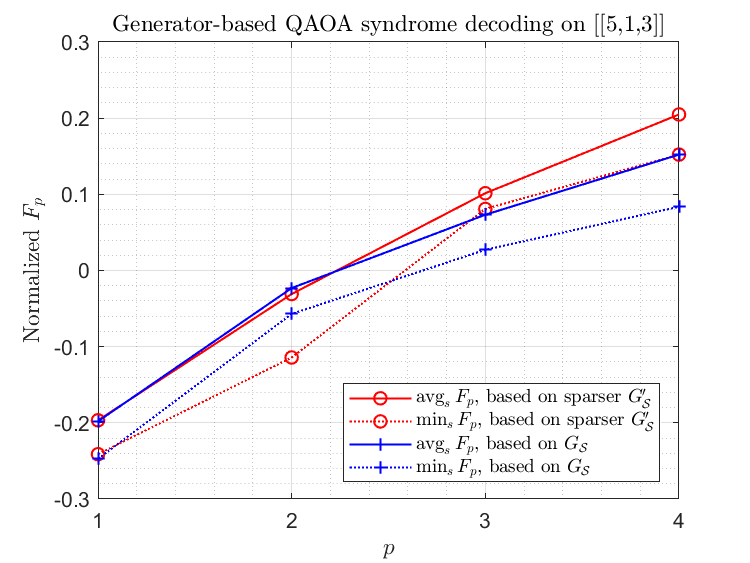}
	\caption{Simulated $F_p$ for  generator-based decoding on the $[[5,1,3]]$ code by QAOA using $G_\cS$ and $G_\cS'$.} \label{fig:513_Fp}		
\end{figure}

The $[[5,1,3]]$ code has been discussed in Example~\ref{ex:513}.
We consider the generator-based decoding Algorithm~\ref{Algorithm:IV}
with the following generator matrices
	\begin{align}
	{G}_{\cS}=\left[\begin{smallarray}{ccccc|ccccc}
 		1 & 0 & 0 & 1 & 0  & 0 & 1 & 1 & 0 & 0\\
 		0 & 1 & 0 & 0 & 1  & 0 & 0 & 1 & 1 & 0\\
 		1 & 0 & 1 & 0 & 0  & 0 & 0 & 0 & 1 & 1\\
 		0 & 1 & 0 & 1 & 0  & 1 & 0 & 0 & 0 & 1\\ 
		\hline
		0 & 0 & 0 & 0 & 0  & 1 & 1 & 1 & 1 & 1\\
		1 & 1 & 1 & 1 & 1  & 0 & 0 & 0 & 0 & 0\\
	\end{smallarray}\right],
	\quad
	{G}_{\cS}'=\left[\begin{smallarray}{ccccc|ccccc}
 		1 & 0 & 0 & 1 & 0  & 0 & 1 & 1 & 0 & 0\\
 		0 & 1 & 0 & 0 & 1  & 0 & 0 & 1 & 1 & 0\\
 		1 & 0 & 1 & 0 & 0  & 0 & 0 & 0 & 1 & 1\\
 		0 & 1 & 0 & 1 & 0  & 1 & 0 & 0 & 0 & 1\\ 
		\hline
				1 & 1 & 0 & 0 & 0  & 0 & 0 & 0 & 1 & 0\\
		0 & 0 & 0 & 1 & 0  & 0 & 0 & 1 & 0 & 1\\
	\end{smallarray}\right]. \label{eq:G_513}
	\end{align}
Note that $G_\cS'$ defines the same code as $G_\cS$ does but $G_\cS'$ has fewer 1's in the last two rows.
	As in the previous subsection, we use NM+basin-hopping and COBYLA+multistart to optimize $F_p$ over  $(\bm\gamma,\bm\beta)\in[0,\pi]^{2p}$ for $p=1,\dots,4$ for different syndromes and the normalized $F_p$ are shown in Fig.~\ref{fig:513_Fp}. {The maximum value of $F_p$ is 5 before normalization.}

The complete QAOA decoding of a quantum code is as follows. 
An  error $\bme\in\{0,1\}^{2n}$ is generated according to an independent depolarizing channel of rate $\epsilon$
with  probability  
	\begin{equation} \label{eq:P_dep}
	P_\text{dep}(\bme) = (\epsilon/3)^{\gw{\bme}} (1-\epsilon)^{n-\gw{\bme}}.
	\end{equation}
If the syndrome of the error is  zero, the decoding output  is $\hat\bme=\bm{0}$.
Otherwise, we simulate the performance of Algorithm~\ref{Algorithm:IV} with better $(\bm\gamma,\bm\beta)$ determined above.
Upon receiving an output distribution together with $T$ possible errors for a given syndrome, 
the error of minimum generalized weight that matches the error syndrome 
will be returned as the output $\hat{\bme}$. 
If there are no errors matching the syndrome, the decoding output is chosen to be $\hat\bme=\bm{0}$.
A decoding (logical) error occurs if and only if $\hat\bme$  is not a degenerate error of $\bme$.
Thus there are $2^{4}$ valid solutions.

\begin{figure}
	\centering \includegraphics[width=0.5\textwidth]{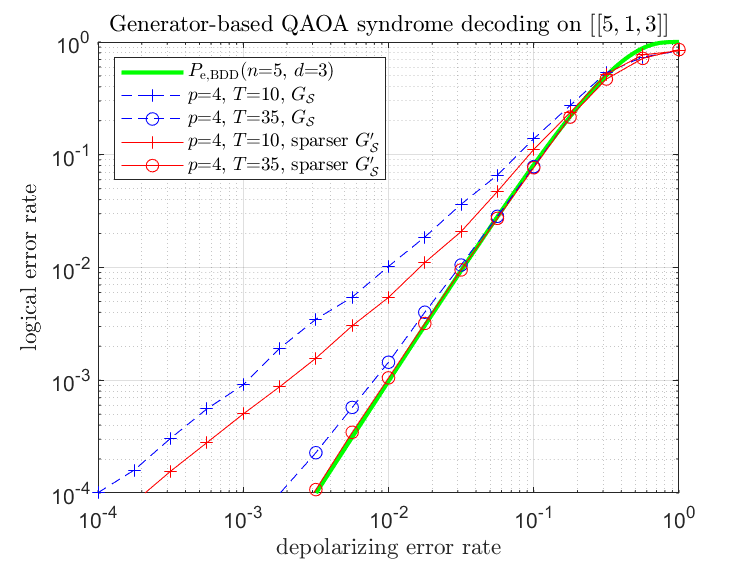}
	\caption{The performance of generator-based syndrome decoding on the $[[5,1,3]]$ code by QAOA.} \label{fig:513_dec}		
\end{figure}

The decoding performance of Algorithm~\ref{Algorithm:IV} is shown in Fig.~\ref{fig:513_dec}.
The $[[5,1,3]]$ code is a perfect code so we compare our QAOA decoding with the BDD performance, which is also plotted ($P_{\rm e,BDD}$ for reference).
As expected, the performance is better using a sparser matrix $G_{\cS}'$. 
For depolarizing rate roughly larger than $0.3$, the logical error rate of QAOA decoder can be better than the BDD performance because of degeneracy.
(See \cite{RW05} and \cite[Fig.~9]{KL21} for more discussions about the performance of $[[5,1,3]]$ in this region.)

\subsubsection{Output distribution by the QAOA decoder} \label{sec:deg_dec}

When the number of qubits in a level-$p$ QAOA is large, classical simulations of all the syndromes are difficult for $p\geq 2$. Thus we may instead  focus on the output distribution of QAOA for a particular syndrome $\bms$ to analyze its resemblance to the real distribution.

Consider again Algorithm~\ref{Algorithm:IV}.
Let $\bmz_\bms\in\{0,1\}^{2n}$  be a vector of syndrome~$\bms$.
Then the occurred error  must be  of the form $\bmu G_\cS + \bmz_\bms$ for some $\bmu\in\{0,1\}^{n+k}$.
The conditional probability of $\bmu$ given $\bms$ is
	\begin{equation} \label{eq:cond_pr}
	P(\bmu|\bms) 
	 = \frac{ P_\text{dep}(\bmu G_\cS+\bmz_\bms) }
			{ \sum_{\bmv\in\{0,1\}^{n+k}} P_\text{dep}(\bmv G_\cS+\bmz_\bms) },
	\end{equation}
where   $P_\text{dep}$ is defined in~\eq{eq:P_dep}.
Let  $Q(\bmu|\bms)$ denote the conditional distribution output by the QAOA  decoder. 
We may evaluate the similarity of $P(\bmu|\bms)$ and $Q(\bmu|\bms)$ using the Jensen--Shannon (JS) divergence \cite{ES03}
$$
J(P\|Q) = \tfrac{1}{2} D(P\|M) + \tfrac{1}{2} D(Q\|M),
$$
where $M=\frac{1}{2}(P+Q)$ and $
D(P\|Q) = \sum_{\bmu\in\{0,1\}^{n+k}} P(\bmu|\bms) \log\frac{P(\bmu|\bms)}{Q(\bmu|\bms)}
$
is the Kullback--Leibler (KL) divergence \cite{KL51}.	
The value of JS divergence is bounded in $[0,1]$ and its value is close to zero for two similar distributions.

\begin{figure*}
	\centering \includegraphics[width=1.0\textwidth]{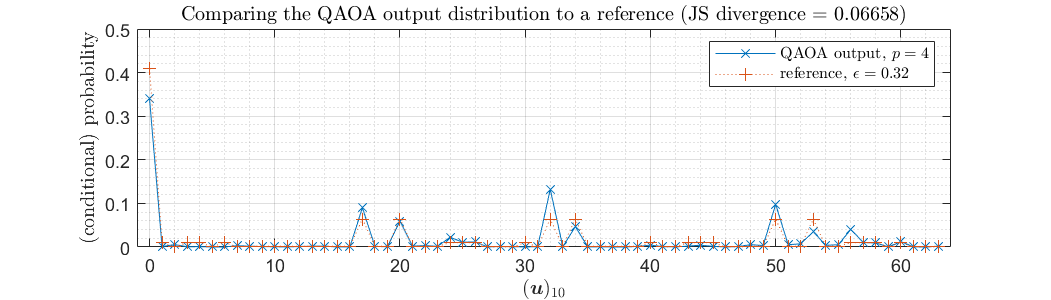}
	\caption{The output distribution of the level-$4$ QAOA decoding on the $[[5,1,3]]$ code by Algorithm~\ref{Algorithm:IV}, given the syndrome of $\hX_1$.} \label{fig:513_pr_X1}		
\end{figure*}

Consider the syndrome of $\hX_1$ (i.e., $\bme = (10000|00000)$)
and we choose $\bmz_\bms=(10000|00000)$.
The approximation by QAOA using the $G'_\cS$ in \eq{eq:G_513} is given  in  Fig.~\ref{fig:513_pr_X1}, with $P$ also plotted for reference.
We find that $J(P\|Q)=0.06658$   at depolarizing rate $\epsilon=0.32$ when $Q$ is the output distribution of the level-$4$ QAOA decoder, 
 as shown in Fig.~\ref{fig:513_pr_X1}.
Note that 
 $$
 (\bmu)_{10} = \textstyle \sum_{i} u_i 2^i
 $$ 
is the decimal representation of $\bmu=(u_1,u_2,\dots,u_{n+k})\in\{0,1\}^{n+k}$. In this case  $\varphi(\hX_1)$  corresponds to $(\bmu)_{10}=0$.

\begin{figure*}
	\centering \includegraphics[width=1.0\textwidth]{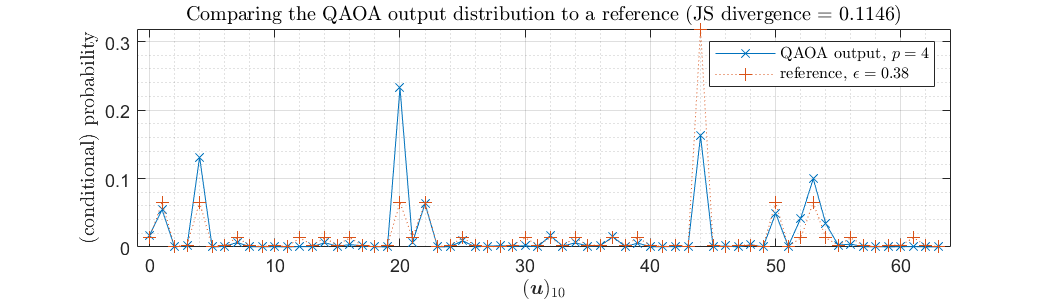}
	\caption{The output distribution of  the level-$4$ QAOA decoding on the $[[5,1,3]]$ code by Algorithm~\ref{Algorithm:IV}, given the syndrome of $\hY_1$.} \label{fig:513_pr_Y1}		
\end{figure*}

Approximating the distribution $P$ is computationally hard. 
Similarly, we have $J(P\|Q)=0.1146$ for   the syndrome of $\hY_1$ at $\epsilon=0.38$ as shown in Fig.~\ref{fig:513_pr_Y1}.
The approximation is not as good as the previous one.


\subsubsection{Shor's $[[9,1,3]]$ code}

Next we study the QAOA decoding on Shor's $[[9,1,3]]$ code \cite{Shor95}, which is a  degenerate code.  
It has a generator matrix \eq{eq:G_S} as
	\begin{align*}
	{G}_{\cS}&=\left[\begin{smallarray}{ccccccccc|ccccccccc}
		0&0&0&0&0&0&0&0&0& 1&1&0&0&0&0&0&0&0\\
		0&0&0&0&0&0&0&0&0& 0&1&1&0&0&0&0&0&0\\
		0&0&0&0&0&0&0&0&0& 0&0&0&1&1&0&0&0&0\\
		0&0&0&0&0&0&0&0&0& 0&0&0&0&1&1&0&0&0\\
		0&0&0&0&0&0&0&0&0& 0&0&0&0&0&0&1&1&0\\
		0&0&0&0&0&0&0&0&0& 0&0&0&0&0&0&0&1&1\\
		1&1&1&1&1&1&0&0&0& 0&0&0&0&0&0&0&0&0\\
		0&0&0&1&1&1&1&1&1& 0&0&0&0&0&0&0&0&0\\
		\hline
		1&1&1&1&1&1&1&1&1& 0&0&0&0&0&0&0&0&0\\
		0&0&0&0&0&0&0&0&0& 1&1&1&1&1&1&1&1&1\\
	\end{smallarray}\right],
	\end{align*}
where the first $n-k$ rows correspond to the generators of its stabilizer group.
To define a reward Hamiltonian for the generator-based decoding problem, we can first perform row operations on the generator matrix 
as long as the rowspace remains unchanged.  Consequently, we  derive the following matrix
	\begin{align*}
	{G}_{\cS}'&=\left[\begin{smallarray}{ccccccccc|ccccccccc}
		0&0&0&0&0&0&0&0&0& 1&1&0&0&0&0&0&0&0\\
		0&0&0&0&0&0&0&0&0& 0&1&1&0&0&0&0&0&0\\
		0&0&0&0&0&0&0&0&0& 0&0&0&1&1&0&0&0&0\\
		0&0&0&0&0&0&0&0&0& 0&0&0&0&1&1&0&0&0\\
		0&0&0&0&0&0&0&0&0& 0&0&0&0&0&0&1&1&0\\
		0&0&0&0&0&0&0&0&0& 0&0&0&0&0&0&0&1&1\\
		1&1&1&0&0&0&0&0&0& 0&0&0&0&0&0&0&0&0\\
		0&0&0&1&1&1&0&0&0& 0&0&0&0&0&0&0&0&0\\
		\hline
		0&0&0&0&0&0&1&1&1& 0&0&0&0&0&0&0&0&0\\
		0&0&0&0&0&0&0&0&0& 0&0&1&0&0&1&0&0&1\\
	\end{smallarray}\right],
	\end{align*}
which is sparser than the original.

\begin{figure*}
	\centering \includegraphics[width=1.0\textwidth]{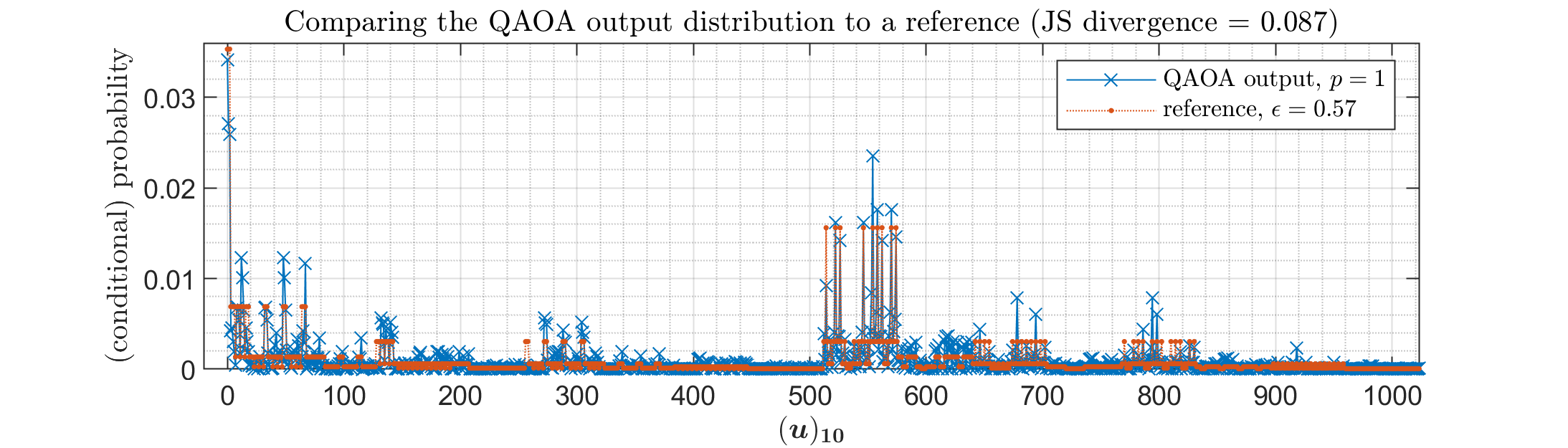}\\
	\includegraphics[width=0.5\textwidth]{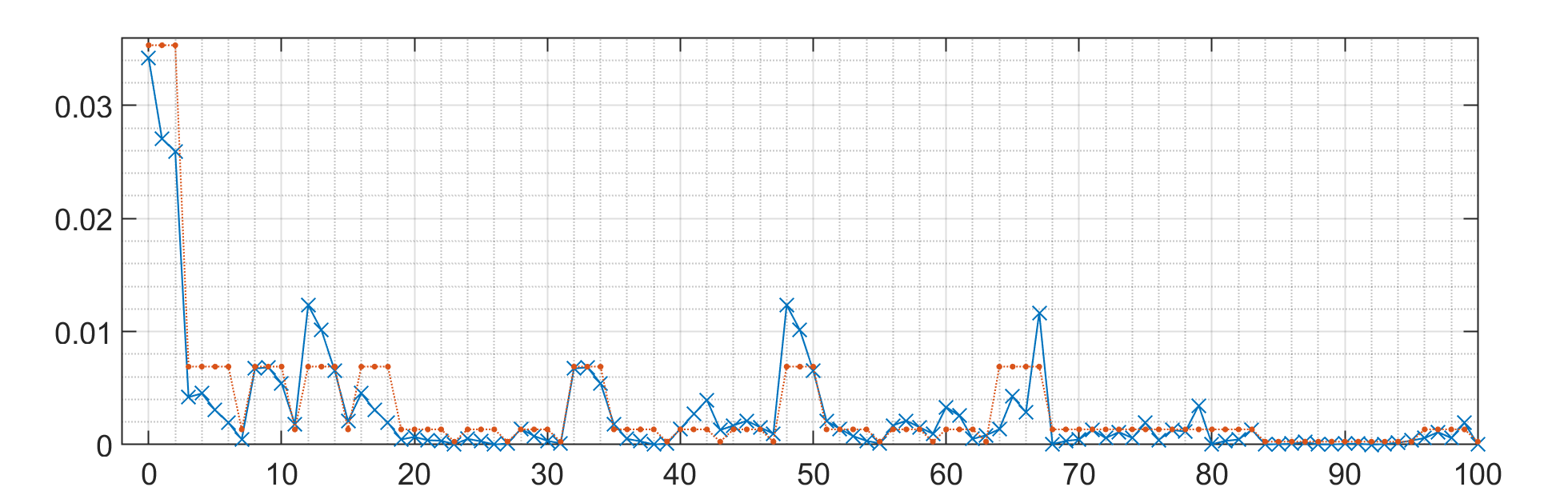}\includegraphics[width=0.5\textwidth]{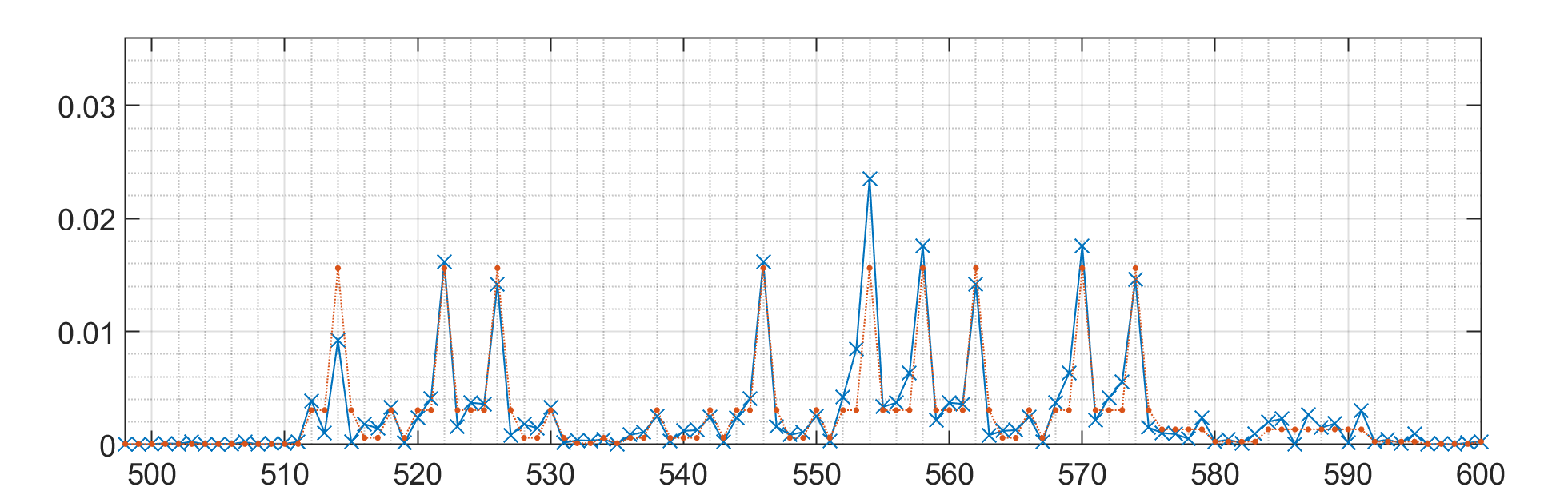}
	\caption{
		The output distribution of the level-$1$ QAOA decoding on the $[[9,1,3]]$ code by Algorithm~\ref{Algorithm:IV}, given the syndrome of $\hZ_2$.
		The regions $(\bmu)_{10}\in[0,100]$ and $(\bmu)_{10}\in[500,600]$ are enlarged in the two lower subfigures.
	} \label{fig:913_p1_Z2}		
	\end{figure*}

We explore the approximation capabilities of QAOA for the conditional distribution \eq{eq:cond_pr} using $G_\cS'$. By considering the syndrome of $\hZ_2$, we demonstrate that the level-1 QAOA decoder based on $G_\cS'$ yields an output distribution $Q$ that closely matches the conditional distribution $P$ at a depolarizing rate of $\epsilon=0.57$ with $J(P|Q)=0.087$, as depicted in Fig.~\ref{fig:913_p1_Z2}.

\begin{figure*}
	\centering \includegraphics[width=1.0\textwidth]{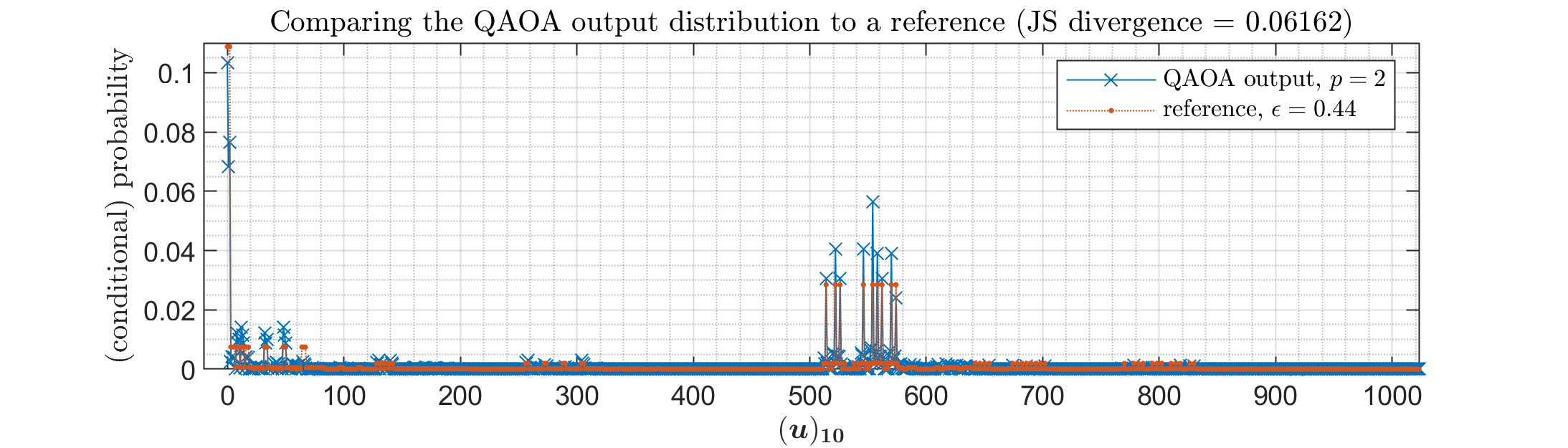}\\
	\includegraphics[width=0.5\textwidth]{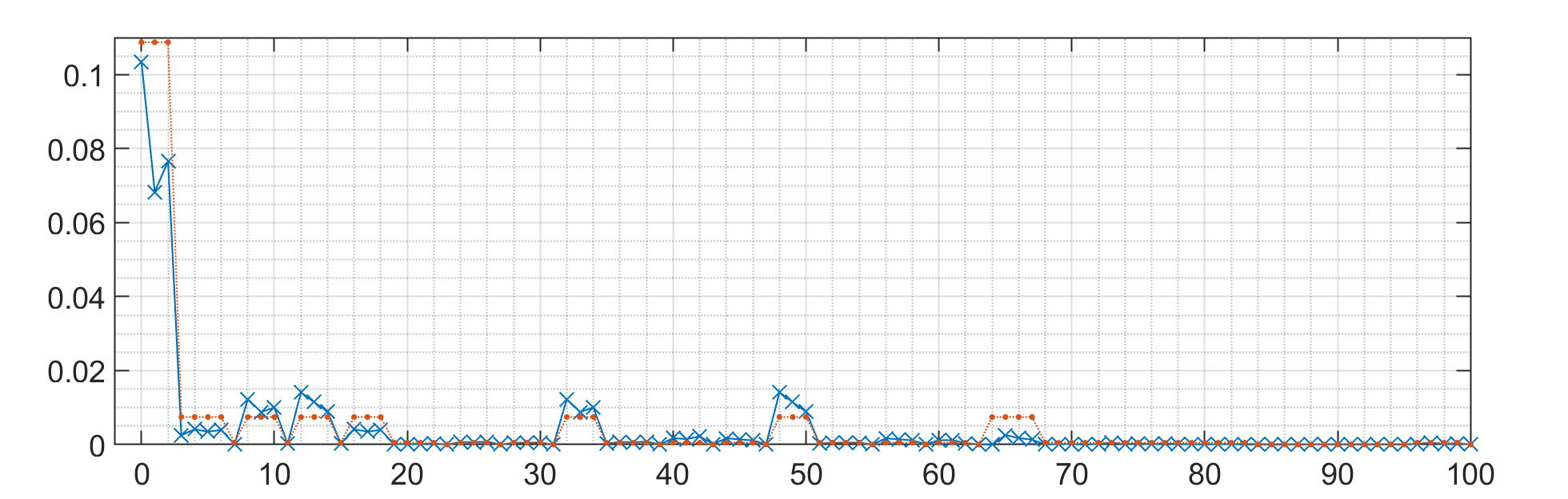}\includegraphics[width=0.5\textwidth]{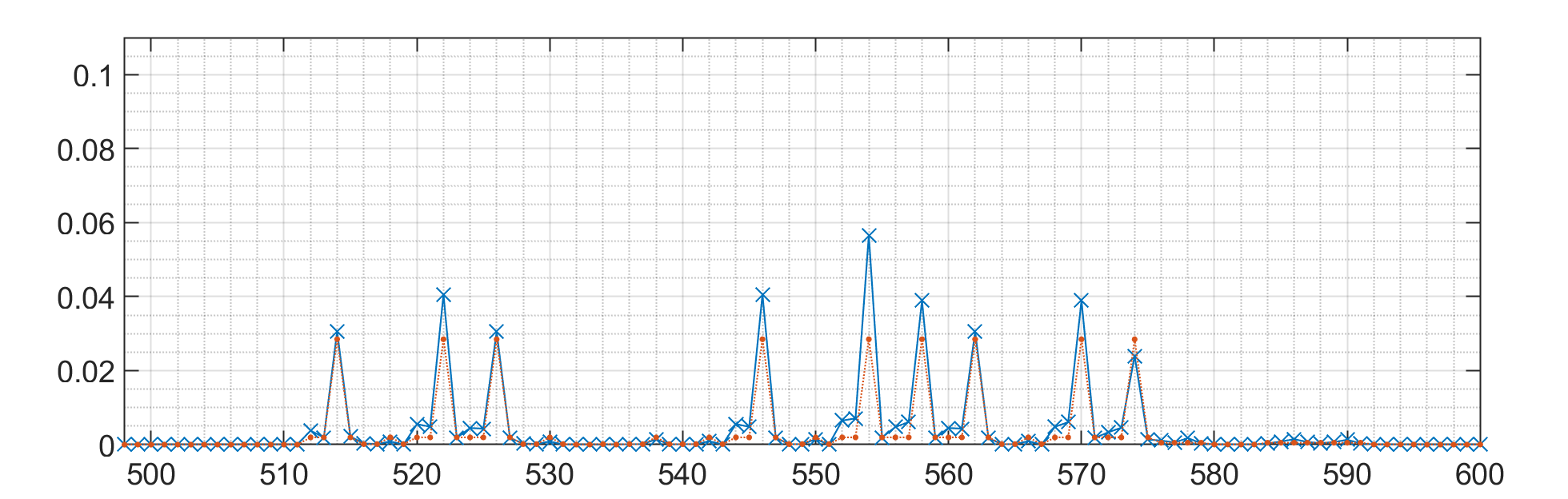}
	\caption{
		The output distribution of the level-$2$ QAOA decoding on the $[[9,1,3]]$ code by Algorithm~\ref{Algorithm:IV}, given the syndrome of $\hZ_2$.
		The regions $(\bmu)_{10}\in[0,100]$ and $(\bmu)_{10}\in[500,600]$ are enlarged in the two lower subfigures.
	} \label{fig:913_p2_Z2}		
	\end{figure*}

Furthermore, by utilizing a higher-level QAOA, we can achieve even smaller values of $J(P|Q)$. For instance, when using the level-2 QAOA decoder with the same syndrome, the output distribution $Q$ closely resembles the conditional distribution $P$ at $\epsilon=0.44$, resulting in $J(P|Q)=0.06162$, as shown in Fig.~\ref{fig:913_p2_Z2}.

It is evident that $\hZ_2$ (corresponding to $(\bmu)_{10}=0$) and its degenerate errors $\hZ_1$ and $\hZ_3$ (corresponding to $(\bmu)_{10}=1$ and~$2$, respectively) possess substantially higher probabilities compared to others. Consequently, the QAOA decoding process is partially degenerate in the sense that it outputs degenerate errors.

  \section{Conclusion}\label{sec:con}

In this paper, we proposed syndrome-based QAOA decoders for both classical and quantum codes, utilizing reward Hamiltonians defined by parity-check or generator matrices.   Additionally, we explored the manipulation of reward Hamiltonian structures to introduce symmetry, thereby defining an equivalent problem that is more amenable to optimization by QAOA.

 For an $[n,k]$ classical code, the generator-induced Hamiltonian requires $k$ qubits, while the check-induced Hamiltonian requires $n$ qubits. On the other hand, for an $[[n,k]]$ quantum code, the generator-induced Hamiltonian necessitates $n+k$ qubits, while the check-induced Hamiltonian requires $2n$ qubits. Consequently, a check-based QAOA decoder may have a higher computational complexity due to the increased number of qubits. However, it offers greater flexibility with two adjustable parameters, $\alpha$ and $\eta$, which control the preference for satisfying checks and error weights.  This adaptability makes the check-based decoder more suited to handle various noise rates.

During our simulations for level $p\le 4$, we observed that the NM+basin-hopping method slightly outperformed the COBYLA+multistart method. Both methods exhibited comparable computational complexities. However, it is essential to note that for $p>4$, the multistart method became significantly slower.

Regarding the performance of QAOA, the problem density is defined as the number of clauses divided by the number of qubits, denoted as $q/m$~\cite{APMB20}. Studies have explored the relationship between problem density and QAOA's performance, revealing that if $q/m>1$, QAOA may exhibit reachability deficits \cite{APMB20}. \bmark{However, in the case of check-based classical decoding using a given $r\times n$ parity-check matrix, the cost Hamiltonian    (\ref{eq:cost_parity})
is defined by $q=r$ constraints and  $m=n$  qubits,
we may assume $r\leq n$ checks are considered 
since there are at most $r\leq n$ independent parity checks. Similarly, in generator-based quantum decoding,  the cost Hamiltonian   (\ref{eq:cost_generator_quantum})
is defined by $q=n$ constraints and   $m=n+k$  qubits.
Therefore, in all our simulations, we always have $q/m\le 1$ in all, which avoids any issues with reachability deficits due to higher problem density caused by additional clauses.}

While implementing a high-level QAOA can be challenging, our  simulations demonstrated promising results in approximating the required distribution for the nine-qubit code, even with just a level-1 QAOA. Notably, QAOA generated a solution distribution in which the probability of each optimal solution was relatively high, suggesting the possibility of finding degenerate errors.

We observed that optimization problems may have multiple optimal solutions, as exemplified by degeneracy in quantum decoding. This characteristic presents the possibility of utilizing QAOA for problems beyond decoding. To estimate each coset probability for degenerate decoding, 
we need to sample the QAOA output multiple times, with each output sample equally weighted to determine the coset probability. However, it is also possible to adjust the weight of each sample, considering factors such as the risk associated with wrong decisions, to serve different purposes. This flexibility opens up potential applications of QAOA in diverse scenarios, such as portfolio optimization \cite{RYXL23}.

 
\rmark{
We used a classical computer to optimize the parameters and simulate the quantum evaluation. This approach limits our current simulations for large code lengths, as the complexity grows exponentially with the number of qubits simulated. In a practical QAOA scenario, the evaluation would be performed by a quantum computer, which would be efficient. The remaining challenge is finding good $(\bm{\gamma},\bm{\beta})$ in $[0,\pi]^{2p}$, requiring an exponentially growing search space with $p$ using an $\varepsilon$-net. Our numerical results on two non-sparse $[7,4,3]$ and $[[5,1,3]]$ codes indicate that the obtained $F_p$ increases at least sublinearly with $p$ when using a limited number of starting points or fixed maximum iterations. This suggests that a bounded classical computation complexity may suffice to obtain a good enough $F_p$ by increasing the level of QAOA.

During the optimization of $F_p$, the classical computer should efficiently compute the expected value of $F_p$ using the quantum computer.  An efficient approach is as follows.
When a scalable quantum computer becomes available, it is crucial to maintain efficient classical computation and classical-quantum interaction. Each time, we obtain an output sample from the quantum computer. To reliably approximate the distribution for evaluating $F_p$, a large but bounded number of samples, such as 10,000, may be required. If the obtained $F_p$ is not high enough, we can increase the level of QAOA. As mentioned previously, $F_p$ is likely to increase at least sublinearly with $p$ using bounded classical search complexity.
When $F_p$  reaches a satisfactory level or when a predefined maximum $p$ is reached, we can stop and collect $T$ samples to find the minimum-weight sample as the answer. During the evaluation and optimization process, every output sample that matches the target syndrome can be buffered. For a code with large 
$n$, if a sample has a weight close to $n\epsilon$, where $\epsilon$ is the physical error rate, it is considered a good answer, and we may choose to stop. Alternatively, if the outputs concentrate in a certain logical coset, we may opt to choose the corresponding logical correction.

Before quantum computers become mature and widely accessible, evaluations using quantum computers may still be expensive. In such cases, Bayesian optimization can be employed to reduce the number of calls to these expensive-to-evaluate functions~\cite{Sha+16,Fra18,Tib+23}.}

\bmark{
Barren plateaus were observed in the simulations, e.g., as can be observed in Fig.~\ref{fig:H743_aXeY} even for $p=1$.
Multistart method or basin-hopping provide the opportunity to escape from a barren-plateau region. 
When the achievable optimum $F_p$ is large, it usually comes with many good local optima, and finding a local optimum is usually sufficient.
For instance, 
Fig.~\ref{fig:H743_a1e4} ($(\alpha,\eta)=(1,4)$) yields an optimum $F_1=1$ and many local optima with $F_1>0.5$, compared to Fig.~\ref{fig:H743_a1e1} ($(\alpha,\eta)=(1,1)$), which has a global optimum $F_1<0.4$.
With multiple starting points or basin-hopping, we easily achieve $F_p>0.5$ with $(\alpha,\eta)=(1,4)$.  
This aligns with the earlier discussion that achieving a good $F_p$
  at a higher level of QAOA may be more beneficial than finding an optimum $F_p$ at a lower level of QAOA.

The barren-plateau effect is theoretically a significant challenge for optimization \cite{Arr+21}. To mitigate this issue, methods to avoid barren plateaus should be considered \cite{Mel+22, Zam+24}.}

\rmark{
For good decoding performance, a high-distance code is required, which introduces high-weight vectors in its generator matrix. If generator-based decoding is needed, a technique that supports good QAOA performance with sparse matrices is necessary. One possible approach is to use a code cover of the original code  \cite{Wib96,For+01,FKV01,KV03}.

The code cover contains pseudo-codewords that are suitable for decoding, from which the original codeword can be inferred. However, using a code cover results in a longer length, thus requiring more qubits. There is a systematic approach to ensure that the length of the code cover is proportional to the original code length, up to a constant factor. For example, a cubic cover has a length of $3n$ if the original code length is $n$. 
Pseudo-codewords tend to have much lower density compared to the original codewords.
Specifically, good regular LDPC codes have a relative distance ($d/n$) that is bounded away from zero, though their relative pseudo-distance may be zero \cite{KV03}. Hence, a sparse generator matrix can potentially be constructed for this code cover to facilitate decoding.}

 \section*{Acknowledgments}
 CYL was supported by the National Science and Technology Council (NSTC) in
 Taiwan, under Grant 111-2628-E-A49-024-MY2, 112-2119-M-A49-007, and 112-2119-M-001-007.

    \bibliographystyle{IEEEtran}



\pagebreak

\appendix

\section{Finding max-cut} \label{sec:maxcut}

\subsection{A graph and its incidence matrix}

The problem of finding max-cut can be approximately solved by QAOA \cite{FGG14a}.
We describe it from the perspective of decoding like what is done in \cite{BB89}.

\bd
A graph is denoted by $\gG = (\gV, \gE)$, where $\gV$ is the set of vertices and $\gE$ is the set of edges.
$\gV$ contains vertices ({\it nodes}) $1,2,\dots,|\gV|$, and 
we denote $\gE = \{\{i,j\}: \text{node $i$ is connected to node $j$ by an edge in $\gG$}\}$.
\ed
\bd
A {\it cut} of $\gG$ is a partition of $\gV$ into two disjoint subsets $\gV_1$ and $\gV_2$.
Let $\gE_1 = { \{\{i,j\}\in\gE: i,j\in \gV_1\} }$ and similarly for $\gE_2$.
Any cut of $\gG$ is corresponding to a {\it cut-set} $\gE' \subseteq \gE$ such that $\gG$ is cut into two disconnected subgraphs 
$\gG_1 = (\gV_1, \gE_1)$ and $\gG_2 = (\gV_2, \gE_2)$ by removing the edges in $\gE'$ from $\gG$.
A {\it max-cut} of $\gG$ is a cut of $\gG$ such that the corresponding cut-set $\gE'$ has maximized $|\gE'|$.
\ed

\be \label{ex:G_633}
Consider a graph $\gG$ of four nodes and six edges as in Fig.~\ref{fig:G_633}.
There are several ways to do the max-cut.
One way is to let $\gV_1 = \{\text{nodes 1 and 4}\}$ (and so $\gV_2 = \{\text{nodes 2 and 3}\}$).
This max-cut corresponds to a cut-set of four edges $\gE'=\{\{1,2\},\{1,3\},\{2,4\},\{4,3\}\}$, which has maximized $|\gE'|=4$ among all possible cut-sets.
	\begin{figure}[h] 
	\centering \resizebox{0.15\textwidth}{!}{\begin{tikzpicture}[node distance=1.3cm,>=stealth,bend angle=45,auto]

\tikzstyle{node}=[circle,draw=black,font=\scriptsize]

\node[node] (v1) at (0,1) {1};
\node[node] (v2) at (2,1) {2};
\node[node] (v3) at (2,0) {3};
\node[node] (v4) at (0,0) {4};

\draw[thin] (v1) -- node[above] {\scriptsize 1} (v2);
\draw[thin] (v2) -- node[right] {\scriptsize 2} (v3);
\draw[thin] (v3) -- node[below] {\scriptsize 3} (v4);
\draw[thin] (v4) -- node[left]  {\scriptsize 4} (v1);
\draw[thin] (v1) -- node[left]  {\scriptsize 5} ++( 1,-0.5) -- (v3);
\draw[thin] (v2) -- node[right] {\scriptsize 6} ++(-1,-0.5) -- (v4);

\end{tikzpicture}}
	\caption{A graph $\gG=(\gV,\gE)$ with $|\gV| = 4$ (nodes) and $|\gE| = 6$ (edges). Edges are labeled with numbers for convenience of description in the main text.} \label{fig:G_633}		
	\end{figure}
\ee

\begin{figure*}
    \centering
    \begin{subfigure}[b]{0.245\textwidth}
        \centering
        \resizebox{0.55\textwidth}{!}{\begin{tikzpicture}[node distance=1.3cm,>=stealth,bend angle=45,auto]

\tikzstyle{node}=[circle,draw=black,font=\scriptsize]

\node[node] (v1) at (0,1) {1};
\node[node] (v2) at (2,1) {2};
\node[node] (v3) at (2,0) {3};
\node[node] (v4) at (0,0) {4};

\draw[thin] (v1) -- (v2);
\draw[thin] (v1) -- (v3);
\draw[thin] (v1) -- (v4);
\draw[thin] (v2) -- (v3);
\draw[thin] (v2) -- (v4);
\draw[thin] (v3) -- (v4);

\end{tikzpicture}}\\~\\
        \!\!\!\!\!\!\includegraphics[width=1.15\textwidth]{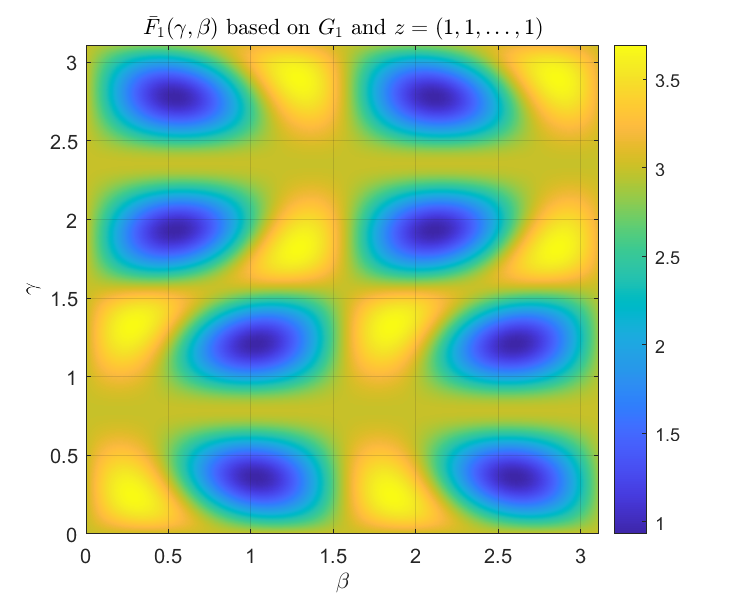}
        \caption{\footnotesize $G_1\in\{0,1\}^{4\times6}$ (a $[6,3]$ code), $\MaxCut=4$ (edges) by cutting $\gV_1=\{\text{nodes 1 and 4}\}$.}	\label{fig:633}
    \end{subfigure}
    \begin{subfigure}[b]{0.245\textwidth}
        \centering
		 \resizebox{0.5\textwidth}{!}{\begin{tikzpicture}[node distance=1.3cm,>=stealth,bend angle=45,auto]

\tikzstyle{node}=[circle,draw=black,font=\scriptsize]

\node[node] (v1) at (0,2) {1};
\node[node] (v2) at (2,2) {2};
\node[node] (v3) at (1,1) {3};
\node[node] (v4) at (0,0) {4};
\node[node] (v5) at (2,0) {5};

\draw[thin] (v1) -- (v4);
\draw[thin] (v1) -- (v3);
\draw[thin] (v1) -- (v2);
\draw[thin] (v2) -- (v3);
\draw[thin] (v2) -- (v5);
\draw[thin] (v3) -- (v4);
\draw[thin] (v3) -- (v5);
\draw[thin] (v4) -- (v5);

\end{tikzpicture}}\\~\\
        \!\!\!\!\includegraphics[width=1.15\textwidth]{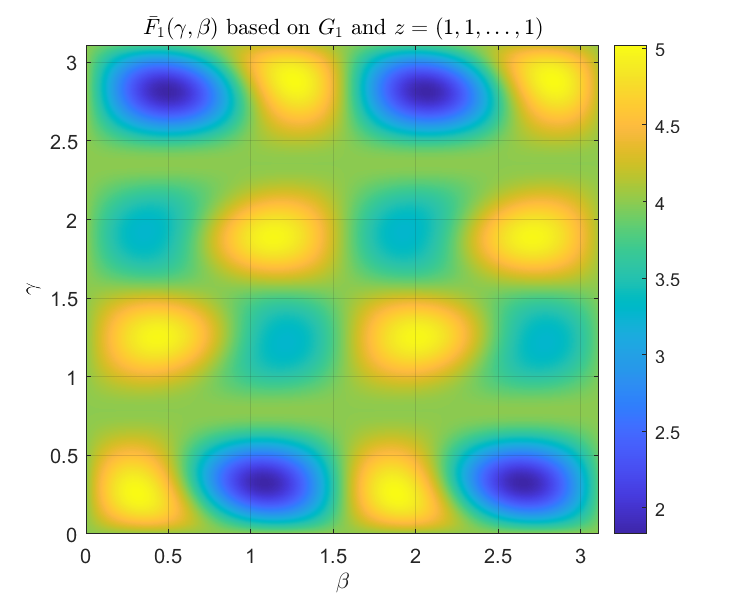}
        \caption{\footnotesize $G_1\in\{0,1\}^{5\times8}$ (an $[8,4]$ code), $\MaxCut=6$ (edges) by cutting $\gV_1=\{\text{nodes 1, 3, 5}\}$.}	 \label{fig:843}
    \end{subfigure}
    \begin{subfigure}[b]{0.245\textwidth}
        \centering
		 \resizebox{0.6\textwidth}{!}{\begin{tikzpicture}[node distance=1.3cm,>=stealth,bend angle=45,auto]

\tikzstyle{node}=[circle,draw=black,font=\scriptsize]

\node[node] (v1) at (1/1.732,2) {1};
\node[node] (v2) at (3/1.732,2) {2};
\node[node] (v3) at (4/1.732,1) {3};
\node[node] (v4) at (3/1.732,0) {4};
\node[node] (v5) at (1/1.732,0) {5};
\node[node] (v6) at (0/1.732,1) {6};

\draw[thin] (v1) -- (v2);
\draw[thin] (v2) -- (v3);
\draw[thin] (v3) -- (v4);
\draw[thin] (v4) -- (v5);
\draw[thin] (v5) -- (v6);
\draw[thin] (v6) -- (v1);
\draw[thin] (v1) -- (v5);
\draw[thin] (v2) -- (v4);
\draw[thin] (v3) -- (v6);

\end{tikzpicture}}\\~\\
        \!\!\!~\includegraphics[width=1.15\textwidth]{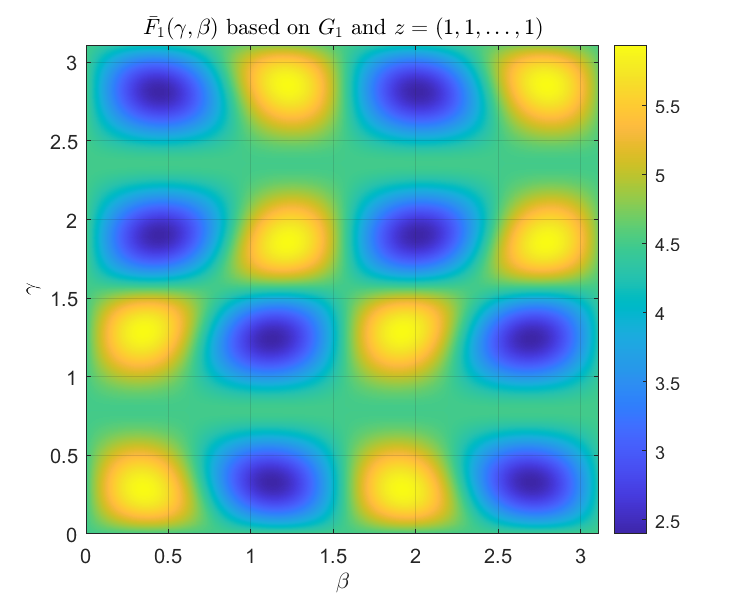}
        \caption{\footnotesize $G_1\in\{0,1\}^{6\times9}$ (a $[9,5]$ code), $\MaxCut=7$ (edges) by cutting $\gV_1=\{\text{nodes 1, 3, 4}\}$.}	\label{fig:953}
    \end{subfigure}
    \begin{subfigure}[b]{0.245\textwidth}
        \centering
		 \resizebox{0.6\textwidth}{!}{\begin{tikzpicture}[node distance=1.3cm,>=stealth,bend angle=45,auto]

\tikzstyle{node}=[circle,draw=black,font=\scriptsize]

\node[node] (v1) at (1/1.732,2) {1};
\node[node] (v2) at (3/1.732,2) {2};
\node[node] (v3) at (4/1.732,1) {3};
\node[node] (v4) at (3/1.732,0) {4};
\node[node] (v5) at (1/1.732,0) {5};
\node[node] (v6) at (0/1.732,1) {6};

\draw[thin] (v1) -- (v2);
\draw[thin] (v2) -- (v3);
\draw[thin] (v3) -- (v4);
\draw[thin] (v4) -- (v5);
\draw[thin] (v5) -- (v6);
\draw[thin] (v6) -- (v1);
\draw[thin] (v1) -- (v4);
\draw[thin] (v2) -- (v5);
\draw[thin] (v3) -- (v6);

\end{tikzpicture}}\\~\\
        \!\!\!~~\includegraphics[width=1.15\textwidth]{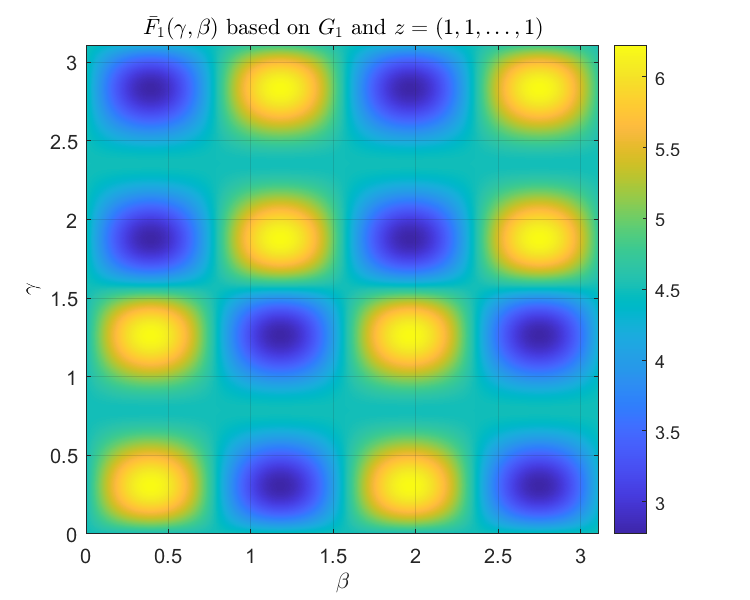}
        \caption{\footnotesize $G_1\in\{0,1\}^{6\times9}$ (a $[9,5]$ code), $\MaxCut=9$ (edges) by cutting $\gV_1=\{\text{nodes 1, 3, 5}\}$.}	\label{fig:953hex}
    \end{subfigure}
    \caption{QAOA ($p=1$) on finding the max-cuts for four graph theoretical codes: for each code, the graph and $\bar F_1(\gamma_1,\beta_1)$ are drawn. 
		It shows that $\max_{(\gamma_1,\beta_1)}\bar F_1(\gamma_1,\beta_1) \ge 0.6924\, \MaxCut$ in each case.}	\label{fig:MaxCut_F1}
\end{figure*}

For a graph $\gG$, label its edges with some order.
For examples, we consider the six edges in Fig.~\ref{fig:G_633} in this order: $\{1,2\},\{2,3\},\{3,4\},\{4,1\},\{1,3\},\{2,4\}$.
That is, $\{1,2\}$ is the first edge, $\{2,3\}$ is the second edge, and so on.

\bd \label{def:inciden_G} 
Give a graph $\gG=(\gV,\gE)$ and an order of edges, an {\it incidence matrix} $G\in\{0,1\}^{|\gV|\times |\gE|}$ is defined by corresponding node $i$ to row $i$ and the $\ell$-th edge to column $\ell$. If the $\ell$-th edge is $\{i,j\}$, then column $\ell$ has ones in entries $i$ and $j$ and has zeros in the other entries.
\ed

Let $\gG$ and $G$ be defined as in Definition~\ref{def:inciden_G}. 
A cut of $\gG$ can be represented by an $\bmu\in\{0,1\}^{|\gV|}$ such that 
$\bmu_i=1$ if and only if node $i$ is in $\gV_1$ (or otherwise, $\bmu_i=0$ and node $i$ is in $\gV_2$).
As shown in \cite{BB89}, a cut $\bmu$ corresponds to a cut-set $\gE'(\bmu)$ that can be represented by a vector $\bmv = \bmu G \in\{0,1\}^{|\gE|}$, 
where $\bmv_\ell=1$ if and only if $\gE'$ contains the $\ell$-th edge of $\gE$.
Also, all $\bmu\in\{0,1\}^{|\gV|}$ correspond to all possible cuts, and 
all possible cut-sets of $\gG$ can be generated by $\bmv = \bmu G$.
A max-cut with maximized $|\gE'(\bmu)|$ corresponds to a $\bmv = \bmu G$ such that $\wt{\bmv}$ is maximized.
Thus, finding a max-cut $\bmu$ with a corresponding maximum cut-set $\gE'(\bmu)$
can be done by
	\begin{align} 
	\argmax_{\bmu\in\{0,1\}^{|\gV|}} |\gE'(\bmu)| &= \argmax_{\bmu\in\{0,1\}^{|\gV|}} \wt{\bmu G} \notag \\ 
		& = \argmin_{\bmu\in\{0,1\}^{|\gV|}} d_H(\bmu G, \bmy),		\label{eq:|E|=wt(v)}
	\end{align}
where $\bmy = (1,1,\dots,1)$.

\be
The graph in Fig.~\ref{fig:G_633} has an incidence matrix
	\begin{equation} \label{eq:G_633}
	G = \left[\begin{smallmatrix}
   1&0&0&1&1&0\\
   1&1&0&0&0&1\\
   0&1&1&0&1&0\\
   0&0&1&1&0&1
	\end{smallmatrix}\right],	
	\end{equation}
%
A maximum-weight vector generated by $G$ is $\bmv = (1,0,1,0,1,1)$, which can be generated by $\bmu=(1,0,0,1)$.
This corresponds to a max-cut with $\gV_1$ containing nodes 1 and 4, as in Example~\ref{ex:G_633}.
The corresponding cut-set $\gE'(\bmu)$ is indicated by $\bmv$ and the number of cut-edges is $|\gE'(\bmu)| = \wt{\bmv}=4$, as also indicated in Example~\ref{ex:G_633}.
\ee

\subsection{The energy topology of QAOA $(p=1)$ for max-cuts}

Given a graph $\gG=(\gV,\gE)$ and its incidence matrix $G$, 
the problem of finding a max-cut can be approximately solved by QAOA by solving \eq{eq:|E|=wt(v)}.
Recall that $G$ has $|\gV|$ rows, but it was known that the rank of $G$ is $|\gV|-1$ \cite{BB89}.
Thus the rowspace of $G$ is an $[n=|\gE|,\,k=|\gV|-1]$ code (called a {\it graph theoretic code} induced by $\gG$).
Since $G$ has one redundant row compared to a generator matrix, we denote it by $G_1$.

Consider that $F_p \in [-r,r]$ is mapped to a non-negative
	\begin{equation} \label{eq:barF_p}
	\bar F_p \triangleq (F_p + r)/2 \in [0,r].
	\end{equation}
Let $\MaxCut = $ the number of cut-edges of a max-cut, i.e.,
	\begin{equation} \label{eq:MaxCut}
	\MaxCut = \max_{\bmu\in\{0,1\}^{|\gV|}} |\gE'(\bmu)|.
	\end{equation}	
Farhi, Goldstone, and Gutmann showed that QAOA can produce a value of $\bar F_p\ge 0.6924\, \MaxCut$ for $p=1$ if the graph $\gG$ is a 3-regular graph \cite{FGG14a}.
We will use several graph theoretic codes to numerically show that, indeed, $\bar F_p\ge 0.6924\, \MaxCut$ for $p=1$, 
and further show that $\bar F_p \to \MaxCut$ for $p\le 4$ for these codes.

We consider four graph theoretic codes defined by the graphs shown in the upper side of Fig.~\ref{fig:MaxCut_F1}.
For an $[n,k]$ code, $G_1$ is a $(k+1)\times n$ matrix.
For generator-based QAOA, $m=k+1$ (which is the number of qubits) and $r=n$ (which is the highest possible value of $\bar F_p$).
Note that the highest value of $\bar F_p$ in this decoding problem is the value of $\MaxCut$, which may be smaller than $n$
because $\bmu G$ may not be able to generate $\bmz = (1,1\dots,1)$. 
For convenience, in each case of Fig.~\ref{fig:MaxCut_F1}, we mention the value of $\MaxCut$ and a max-cut to reach it in the subcaption.
To see how QAOA can approximate the solution, we run Algorithm~\ref{Algorithm:II} by assigning $G=G_1$ and $\bmz = (1,1,\dots,1)$ in step~1). The results of $p=1$ is shown in the lower side of Fig.~\ref{fig:MaxCut_F1}, by drawing $\bar F_1$ as an energy topology over $(\gamma_1,\beta_1)\in[0,\pi]\times[0,\pi]$.
For the first three cases, QAOA can produce a value of $\bar F_1$ much larger than $0.6924\, \MaxCut$.
Note that the second case is not a 3-regular graph (since node~3 has degree~4) but QAOA can still produce a high value of $\bar F_1$.
For the last case, the value of $\MaxCut$ is $n$, and QAOA can produce a value of $\bar F_1 \approx 0.6924\, \MaxCut$.

We should discuss the use of redundant rows before we proceed to the case of larger $p$.
A generator matrix $G$ by definition is without redundant rows.
It is possible to remove the redundant row in the incidence matrix $G_1$. 
This allows to use fewer qubits in QAOA.
However, then the generator matrix does not directly reflect the structure of the graph $\gG$.
In the energy topology such as in Fig.~\ref{fig:633}, maximum $\bar F_1$ becomes slightly higher, but for $\beta\in [1,2]$, the high $\bar F_1$ regions disappear. 
That is, there are fewer good $(\gamma,\beta)$ points to target to.
This would make the optimization harder when $p$ increases, and thus we use incidence matrix $G_1$ to proceed.

\subsection{Finding max-cuts by QAOA $(p\ge 1)$}

Next, we increase $p$ to make $\bar F_p \to \MaxCut$ for each case in Fig.~\ref{fig:MaxCut_F1}.
A set of good parameters $(\bm\gamma,\bm\beta)\in [0,\pi]^p\times [0,\pi]^p$ is needed for QAOA to perform well to have a large $F_p(\bm\gamma,\bm\beta)$,
and finding such $(\bm\gamma,\bm\beta)$ is an optimization problem.
To have lower complexity, derivative-free methods are more suitable for this optimization problem \cite{SSL19}.
We do the optimization by using {\tt SciPy} \cite{Vir+20}, which can efficiently perform derivative-free methods such as the {\it Nelder-Mead} (NM) method \cite{NM65}
and {\it constrained optimization by linear approximation} (COBYLA) \cite{Pow78,Pow94}. 
Since the problem is not convex, we also need some schemes to prevent that bad parameters are chosen.
We run multiple NM or COBYLA with different starting points $(\bm\gamma,\bm\beta)$ and select the best result, 
called the {\it multistart} method (see \cite{SSL19} for several ways to improve the multistart efficiency). 
We also run NM or COBYLA with {\it basin-hopping} \cite{WD97} to do a heuristic optimization. 

The optimization procedure is sensitive to the starting point, especially when $p$ increase, since $(\bm\gamma,\bm\beta)$ is in a large space with dimension $2p$.
In \cite[Sec.~VI]{FGG14a}, it is suggested to use small $\gamma$ and $\beta$ with large $p$.
The results in Fig.~\ref{fig:MaxCut_F1} support this argument even for $p=1$. 
This is because we want QAOA to carefully evolve with some control.
By taking $\gamma$ as an example, observe~Eqs.~(\ref{eq:C_j})--(\ref{eq:UC}): 
QAOA assigns larger $\hat C_j$ for preferred $\ket{\bmx}$ and makes $e^{-\gamma\hat C_j}$ contained in the evolution. 
The effective range of angular velocity is $[-\pi,\pi]$, inherent from $C_j\in\{-1,+1\}$.
Define $[\gamma\hat C_j]_{\pm\pi}\in[-\pi,\pi]$ as the remainder of $\gamma\hat C_j$ divided by~$\pi$.
A small $\gamma$ guarantees that if $\hat C_{j_1} < \hat C_{j_2} < \dots < \hat C_{j_r}$ then there is a similar order
$[\gamma \hat C_{j_1}]_{\pm\pi} < [\gamma \hat C_{j_2}]_{\pm\pi} < \dots < [\gamma \hat C_{j_r}]_{\pm\pi}$ after taking the remainder. 
Of course, we also need not-too-small $\gamma$ to have effective evolution. 

\begin{figure}
	\centering \includegraphics[width=0.5\textwidth]{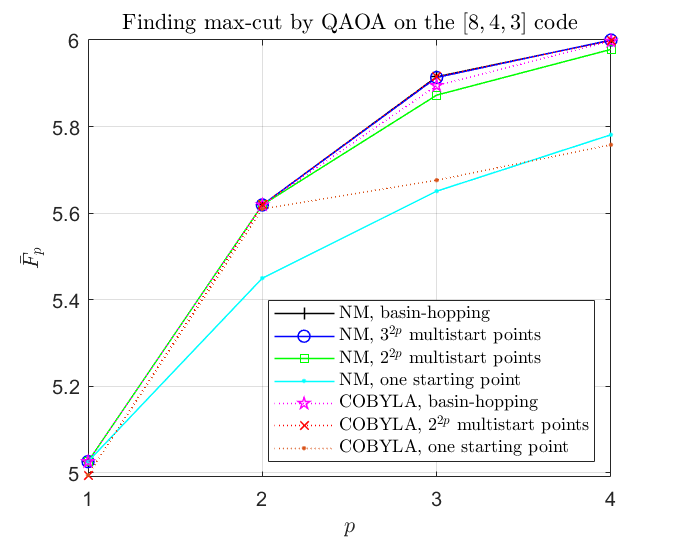}
	\caption{Comparing several optimization procedures by using the $[8,4,3]$ code in Fig.~\ref{fig:843} ($\MaxCut=6$).} \label{fig:843_maxcut}
	\end{figure}

We do some investigation by using the $[8,4,3]$ code in Fig.~\ref{fig:843}, since its graph is not 3-regular and QAOA is more sensitive to bad parameters.
Indeed, we find that, if we let the basin-hopping start from $(\gamma_p,\beta_p)\approx(0,0)$ for all $p$, then both the NM method and COBYLA perform very well and almost the same performance.
If we adjust the starting point to $(\gamma_p,\beta_p)=(1,1)$ for all $p$, then NM performs the same well but COBYLA performs slightly worse. 
This is showed in Fig.~\ref{fig:843_maxcut}.

On the other hand, the multistart method uses multiple optimization instances with different starting points to prevent the problem of bad starting points. 
In the multistart method, we find $\kappa\ge 2$ points equally separated on $[0,\pi]$ and this defines $\kappa^{2p}$ starting points in $[0,\pi]^p\times[0,\pi]^p$.
The complexity increases exponentially with $p$, even for the smallest $\kappa=2$.
Increasing $\kappa>2$ can let the multistart method approach the optimal solution, but this further increases the complexity $(\frac{\kappa}{2})^{2p}$ times.
Some cases of $\kappa=2$ or 3 are also in Fig.~\ref{fig:843_maxcut}.

We remark that the complexity of the multistart method can be reduced by sharing the evaluations of $F_p(\bm\gamma,\bm\beta)$ 
for the same $(\bm\gamma,\bm\beta)$ required between different optimization instances \cite{SSL19}. 

For reference, we also provide the results of ``one starting point'' (i.e., without basin-hopping or multistart) in Fig.~\ref{fig:843_maxcut}, and as expected, the performance is not good.

\begin{remark} \label{rm:all_maxcut}
Unlike a classical algorithm that only a solution will be returned, QAOA can find all possible solutions. 
For example, there are four ways of choosing $\gV_1$ in Fig.~\ref{fig:843} to form a max-cut.
In Fig.~\ref{fig:843_maxcut}, when QAOA has $\bar F_p$ achieving $\MaxCut=6$, 
its output has $\ket{\bmx}$ with $\bmx =$ (0,1,0,1,0), (0,1,1,1,0), (1,0,0,0,1) or (1,0,1,0,1) in equal probability 0.25.
(The solution (1,0,1,0,1) corresponds to $\gV_1 = \{\text{nodes 1, 3, 5}\}$ indicated in Fig.~\ref{fig:843}.)
By running QAOA for multiple times $T$ as in Algorithm~\ref{Algorithm:I}, all the solutions can be obtained.
\end{remark}

\begin{figure}[h] 
	\centering \includegraphics[width=0.5\textwidth]{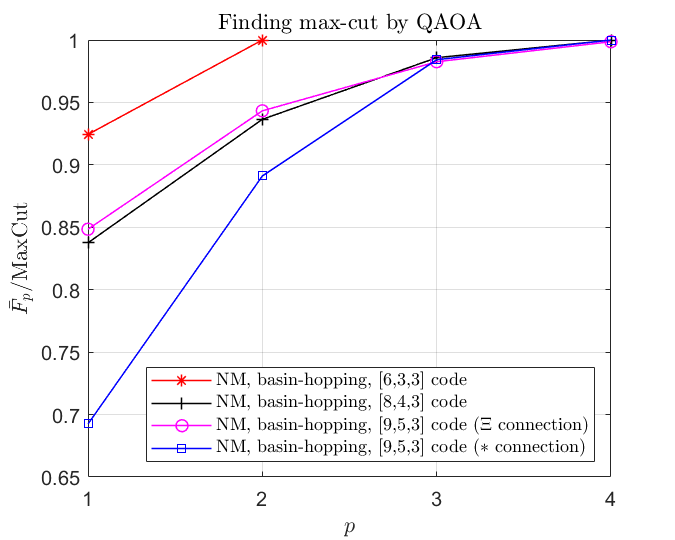}
	\caption{QAOA performance for finding the max-cuts (largest-weight codewords) of the codes in Fig.~\ref{fig:MaxCut_F1}.} \label{fig:maxcuts} 
	\end{figure}

Finally, we use the NM method with basin-hopping to find proper parameters for each case in Fig.~\ref{fig:MaxCut_F1} and show that QAOA have $\bar F_p \to \MaxCut$ for each case. The each code has a minimum distance three. To differentiate the third and fourth cases (where both are $[9,5,3]$ codes), the code in Fig.~\ref{fig:953} is denoted by ``$[9,5,3]$ ($\Xi$ connection)'' and the code in Fig.~\ref{fig:953hex} is denoted by ``$[9,5,3]$ ($*$ connection)''.
The results are shown in Fig.~\ref{fig:maxcuts}.

\end{document}